\documentclass[graybox]{svmult}

\usepackage{mathptmx}      
\usepackage{helvet}        
\usepackage{courier}       
\usepackage{type1cm}       
\usepackage{makeidx}         
\usepackage{graphicx}       
\usepackage{multicol}        
\usepackage[bottom]{footmisc}

\usepackage{amsmath}
\usepackage{amssymb}
\usepackage{amsfonts}

\usepackage{amsthm}

\newcommand{\mcc}{\mathcal{C}}
\newcommand{\mcd}{\mathcal{D}}
\newcommand{\mcl}{\mathcal{L}}
\newcommand{\mcm}{\mathcal{M}}
\newcommand{\mcp}{\mathcal{P}}
\newcommand{\mbc}{\mathbb{C}}
\newcommand{\mbh}{\mathbb{H}}
\newcommand{\mbr}{\mathbb{R}}

\newcommand{\mfa}{\mathfrak{a}}
\newcommand{\mfb}{\mathfrak{b}}

\newcommand{\Res}{\text{Re}(s)}
\newcommand{\tab}{\hspace{5mm}}
\newtheorem{conj}[theorem]{Conjecture}
\newtheorem{cor}[theorem]{Corollary}
\newtheorem{rem}[theorem]{Remark}

\begin{document}
\title*{The Sound of Fractal Strings and the Riemann Hypothesis}
\author{Michel L. Lapidus}
\institute{University of California \\Department of Mathematics \\900 University Avenue\\ Riverside, CA 92521--0135, USA\\ \email{lapidus@math.ucr.edu}
}
%
%

\maketitle

\abstract{We give an overview of the intimate connections between natural direct and inverse spectral problems for fractal strings, on the one hand, and the Riemann zeta function and the Riemann hypothesis, on the other hand (in joint works of the author with Carl Pomerance and Helmut Maier, respectively). We also briefly discuss closely related developments, including the theory of (fractal) complex dimensions (by the author and many of his collaborators, including especially Machiel van Frankenhuijsen), quantized number theory and the spectral operator (jointly with Hafedh Herichi), and some other works of the author (and several of his collaborators).}

\vspace{2cm}


\vspace{1cm}

\noindent Suggested short title for the paper: "Fractal Strings and the Riemann Hypothesis."

\tableofcontents

\section{Riemann Zeros and Spectra of Fractal Strings: An Informal Introduction}\label{S1}

Unlike an ordinary (Sturm--Liouville) vibrating string, which consists of a single interval (of length $\ell$, say), a fractal string consists of infinitely many intervals (of lengths $\ell_1, \ell_2, \cdots, \ell_j, \cdots,$ with $\ell_j \downarrow 0$ as $j \rightarrow \infty$), vibrating independently of each other. Hence, the (eigenvalue or frequency) spectrum of a fractal string consists of the union of the spectra (counting multiplicities) of each of the countably many ordinary strings of which it is composed. \\

\tab A fractal string (or, equivalently, its boundary, viewed as a compact subset of the real line $\mathbb{R}$) always has (fractal) Minkowski dimension $D$ between $0$ and $1$, the most extreme case $D=0$ and $D=1$ being referred to (following [Lap1]) as the least and most fractal case, respectively, while the case when $D=1/2$ is referred to (also as in [Lap1]) as the {\em midfractal case.} The latter case will play a key role throughout this paper. \\

\tab By listening to a fractal string, one can detect whether or not one of its complex dimensions coincides with a nontrivial Riemann zero (that is, with a zero of the Riemann zeta function which is located in the critical strip $0 < \Res < 1$). Indeed, it turns out that the Riemann zeta function $\zeta = \zeta (s)$ mediates between the geometry and the spectrum of a fractal string: 

\begin{equation}\label{1.1}
\zeta_\nu (s) = \zeta(s) \cdot \zeta_\mcl(s),
\end{equation}
where $\zeta_{\nu, \mcl} (s) = \zeta_\nu (s) := \sum_{k=1}^\infty f_k^{-s}$ {\em is the spectral zeta function} of the fractal string $\mcl$,  with $\{f_k \}_{k=1}^\infty$ denoting the sequence of (suitably normalized) frequencies of $\mcl$, written in nondecreasing order according to their multiplicities, and $\zeta_\mcl (s) := \sum_{j=1}^\infty \ell_j^s$ is the {\em geometric zeta function} of $\mcl$. This relation (discovered in \cite{Lap2, Lap3}) has played an important role in fractal string theory, as it enables one to understand why some of the complex dimensions of $\mcl$ (i.e., the poles of the meromorphic continuation of $\zeta_\mcl$) may be ``canceled'' by (nontrivial) zeros of 
$\zeta$. Accordingly, certain oscillations which are present in the intrinsic  geometry of the fractal string are no longer ``visible'' (or rather, ``audible'') in the spectrum of $\mcl$. \\

\tab More poetically, it is shown in [LapMai1,2] that ``{\em One can hear the shape of a fractal string of dimension} $D \neq 1/2$'' (in the sense of a certain inverse spectral problem, denoted by (ISP)$_D$ and to be specified in \S6 below, and not in the original sense of Mark Kac \cite{Kac}) if and only if the Riemann hypothesis is true. Moreover, one cannot hear it for all of the fractal strings of dimension $1/2$ (because  $\zeta = \zeta (s)$ has zeros on the critical line $\Res = 1/2$). Hence, in the present approach, the truth of the Riemann hypothesis is equivalent to the existence of a (mathematical) phase transition at $D = 1/2$ and at no other dimension $D$ in the ``critical interval'' $(0,1)$. (See Theorems \ref{T6.2} and \ref{T6.4} below.) By ``hearing the shape of a fractal string'' here, we mean that the inverse spectral problem (ISP)$_D$ in \S6 below has an affirmative answer for any fractal string of dimension $D$.\\

\tab Phrased a little more precisely: one can hear the shape of a fractal 
string of a given (Minkowski) dimension $D \in (0,1)$ if and only if the Riemann zeta function $\zeta (s)$ does not have any zero on the vertical line 
$\Res = D$. (See Theorem \ref{T6.1}.) Furthermore, in general, one cannot hear the shape of a fractal string in the midfractal case where $D=1/2$. (See Corollary \ref{C6.3}.) Consequently, one can hear the shape of a fractal string in every possible dimension $D \in (0,1)$ (other than $1/2$) if and only if the Riemann hypothesis \cite{Rie} is true; that is, if and only if 

\begin{equation}\label{2.1}
\zeta(s) = 0, \quad 0 < \Res < 1 \Rightarrow \Res = \frac{1}{2}.
\end{equation}
(See Theorem \ref{T6.2}.) These results have been established by the author and Helmut Maier in \cite{LapMai2} (announced in 1991 in \cite{LapMai1}) building on the author's earlier work \cite{Lap1} (see also [Lap2,3]) on a partial resolution of the Weyl--Berry conjecture in any dimension [Berr1,2]) as well as on the ensuing work of the author and Carl Pomerance \cite{LapPom2} (announced in 1990 in \cite{LapPom1}) on a resolution of the one-dimensional (modified) Weyl--Berry conjecture (of \cite{Lap1}) and its unexpected connections with the Riemann zeta function. (See Conjecture \ref{C5.1} and Theorem \ref{T5.3} in \S5.)\\

\tab Later on, these results of \cite{LapMai2} (which made use at the heuristic level of the intuition of complex dimensions) were reinterpreted (by the author and Machiel van Frankenhuijsen in [Lap-vFr1,2]) in terms of the then rigorously defined notion of complex dimension, as well as extended to a large class of Dirichlet series and integrals, including all of the arithmetic zeta functions for which the generalized Riemann  hypothesis is expected to hold. (See also [Lap-vFr3, Chapter 9].) Moreover, a method similar to the one used in \cite{LapMai2}, but now relying in part on the explicit formulas established in [Lap-vFr1--3] (and direct computations as well as on inverse spectral problems), was used to show that the Riemann zeta function, along with a large class of Dirichlet series and integrals (including most of the arithmetic zeta functions, [ParSh1--2], [Sarn], [Lap-vFr3, Appendix A] or [Lap6, Appendices B, C and E], other than the zeta functions of varieties over finite fields, for which the result clearly does not hold), cannot have an infinite vertical arithmetic progression of 
zeros. (See Lap-vFr3, Chapter 11] for this result and several extensions concerning the density of the zeros.) \\

\tab Unknown at the time to the authors of \cite{Lap-vFr1} (and of earlier papers on this and related subjects), this latter result about the zeros in arithmetic progression, in the special case of $\zeta$, was already obtained by Putnam in [Put1--2] by a completely different method, which could not be generalized to this significantly broader setting. This turned out to be quite beneficial to the general theory of complex dimensions as it led us in [Lap-vFr1--3] to significantly improve and refine the authors' original pointwise and distributional explicit formulas. (See, e.g., [Lap-vFr3, Chapter 5].)\\

\tab In the rest of this paper, we will be more specific and explain what type of inverse spectral problem is involved here. (See, especially, \S6.) First of all, we will need to precisely  define (in \S2) what is a fractal string as well as its Minkowski dimension and content.\\

\tab We will then recall (in \S3) results from [LapPom1,2] providing a characterization of the notions of Minkowski measurability and nondegeneracy, which will play a key role in \S5 and part of \S6 (as well as serve as a motivation for some aspects of \S 7.4). In \S4, we will discuss Weyl's asymptotic formula and conjecture [We1,2] for the spectral asymptotics of drums with smooth (or sufficiently ``regular'') boundary, as well as the Weyl--Berry conjecture [Berr1,2] for drums with fractal boundary (or ``fractal drums'') and its partial resolution obtained in \cite{Lap1}. In \S5, we will present the resolution of the modified Weyl--Berry (MWB) conjecture \cite{Lap1} for fractal strings obtained in [LapPom1,2], thereby establishing a precise connection between the corresponding {\em direct spectral problem} and the Riemann zeta function $\zeta = \zeta(s)$ in the critical interval $0 < s < 1$. In \S6, we will introduce the aforementioned  {\em inverse spectral problem} (ISP)$_D$, for each $D \in (0,1)$, and precisely state the results of [LapMai1,2] connecting it with the presence of zeros of the Riemann zeta function in the critical strip $0 < \Res < 1$ (the so-called critical or nontrivial zeros), and thereby, with the Riemann hypothesis.\\

\tab Finally, in \S7, we will discuss a variety of topics, closely connected to (or motivated in part by) the above developments. The subjects to be discussed include the mathematical theory of complex dimensions of fractal strings developed in [Lap-vFr1--3] (see \S 7.2), its higher-dimensional counterpart recently developed in \cite{LapRaZu1} and [LapRa\u Zu2--8] (see \S 7.3), as well as aspects of ``quantized number theory'' (see \S 7.4) developed in \cite{HerLap1} (and [HerLap2--4], along with \cite{Lap7}). In particular, in the latter subsection, we will see that the aforementioned inverse spectral problem (ISP)$_D$ can be rigorously reinterpreted in terms of the invertibility (or ``quasi-invertibility'') of the ``spectral operator'' $\mfa = \zeta (\partial)$, with the ``infinitesimal shift'' (of the real line) $\partial$ now playing the role of the usual complex variable $s$ in the definition of the quantum (or operator-valued) analog of the classic Riemann zeta function $\zeta = \zeta(s)$.\\

\tab We close this introduction by providing several relevant references. For general references concerning  the theory of the Riemann zeta  function and related aspects  of analytic number theory, we mention, for example, [Edw, Ing, Ivi, KarVor, Lap6, Lap-vFr3, ParSh1--2, Pat, Ser, Ti] along with the relevant references therein. For fractal string theory and the associated theory of complex dimensions, along with their applications to a variety of subjects, including fractal geometry, spectral geometry, number theory and dynamical systems, we refer to \cite{Lap-vFr3}, along with [EllLapMaRo, HamLap, Fal2, HeLap, HerLap1--5, LalLap1--2, Lap1--9, LapL\' eRo, LapLu1--3, LapLu-vFr1--2, LapMai1--2, LapNe, LapPe1--3, LapPeWi1--2, LapPom1--3, LapRa\u Zu1--8, LapRo1--2, LapRo\u Zu, L\' eMen, MorSepVi, Pe, PeWi, Ra, RatWi2, Tep1--2, \u Zu1--2] and the relevant references therein. In particular, Chapter 13 of \cite{Lap-vFr3} provides an exposition of a number of recent extensions and applications of the theory, including to fractal sprays (higher-dimensional analogs of fractal strings, \cite{LapPom3}) and self-similar systems ([Lap-vFr3, \S 13.1], based on [LapPe2--3, LapPeWi1--2, Pe, PeWi]), $p$-adic (or nonarchimedean) geometry, ([Lap-vFr3, \S 13.2], based on [LapLu1--3, LapLu-vFr1--2]), multifractals ([Lap-vFr3, \S 13.3], based on [LapRo, LapL\' eRo, EllLapMaRo]), random fractal strings ([Lap-vFr3, \S 13.4], based on [HamLap]), as well as fractal membranes and the Riemann (or modular) flow on the moduli space of fractal membranes ([Lap-vFr3, \S 13.5], based on the book \cite{Lap6} and on \cite{LapNe}). As a general rule, we will give specific references to the most recent monograph \cite{Lap-vFr3} rather than to the earlier research monographs, \cite{Lap-vFr1} and \cite{Lap-vFr2}.

\section{Fractal Strings and Minkowski Dimension}\label{S2}
A (nontrivial) fractal string $\mcl$ is a bounded open set $\Omega \subset \mathbb{R}$ which is not a finite union of intervals. Hence, $\Omega$ is an infinite countable disjoint union of (bounded) intervals $I_j,$ of lengths $\ell_j = |I_j|$, for $j = 1,2, \cdots: \Omega = \cup_{j=1}^\infty I_j$. Since $|\Omega| = \sum_{j=1}^\infty \ell_j < \infty$, we may assume without loss of generality that $(\ell_j)_{j=1}^\infty$ is nonincreasing and so $\ell_j \downarrow 0$ as $j \rightarrow \infty$. Furthermore, for our purposes, we may identify a fractal string with its associated sequence of lengths (or scales), $\mcl := (\ell_j)_{j=1}^\infty$, written as above (counting multiplicities). Indeed, all of the geometric notions we will work with, such as $V(\varepsilon)$, the Minkowski  dimension and content, as well as the geometric zeta function and the complex dimensions, depend only on $\mcl = (\ell_j)_{j=1}^\infty$ and not on the particular geometric realization of $\mcl$ as a bounded open subset $\Omega$ of $\mathbb{R}$. \\

\tab Given $\varepsilon > 0$, the (inner) $\varepsilon$-{\em neighborhood} (or {\em inner tube}) of $\partial \Omega$ (or of $\mcl$) is given by 

\begin{equation}\label{2.2}
\Omega_\varepsilon := \{x \in \Omega : d(x, \partial \Omega) < \varepsilon \},
\end{equation}
where $\partial \Omega$ denotes the boundary of $\Omega$ (a compact subset of $\mathbb{R}$) and $d(\cdot, \partial \Omega)$ denotes the (Euclidean) distance to $\partial \Omega$.\\

\tab In the sequel, we will let 

\begin{equation}\label{2.3}
V(\varepsilon) = V_\mcl (\varepsilon) := |\Omega_\varepsilon|,
\end{equation}
the volume (really, the length or 1-dimensional Lebesgue measure) of $\Omega_\varepsilon$. As was mentioned above, it can be shown (cf. [LapPom1,2]) that $V(\varepsilon)$ depends only on $\mcl$ (and not on the particular geometric realization $\Omega$ of $\mcl$). Then, given $d \geq 0$, the $d$-{\em dimensional upper Minkowski content} of $\mcl$ (or of $\partial \Omega$) is given by 

\begin{equation}\label{2.4}
\mcm_d^* = \mcm_d^* (\mcl) := \lim_{\varepsilon \rightarrow 0^+} \sup \frac{V(\varepsilon)}{\varepsilon^{1-d}},
\end{equation}
and similarly for the $d$-dimensional lower Minkowski content $\mcm_{*,d} = \mcm_{*,d}(\mcl)$, except for the upper limit replaced by a lower limit.\\

\tab The {\em Minkowski dimension} of $\mcl$ (or of its boundary $\partial 
\Omega$),\footnote{This is really the upper Minkowski dimension of $\partial \Omega$, relative to $\Omega$, but we will not stress this point in this paper (except perhaps in \S 7.3).} 
denoted by $D = D_\mcl$, is defined by 

\begin{align}\label{2.5}
D &:= \inf \{d \geq 0 : \mcm_d^* < \infty \}\\
\notag &= \sup \{ d \geq 0: \mcm_d^* = + \infty \}.
\end{align}
We note that since $|\Omega| < \infty$, we always have $0 \leq D \leq 1$. Furthermore, from the physical point of view, $D$ will play the role of a {\em critical exponent} (or {\em critical parameter}): it is the unique real number $D$ such that $\mcm_d^* = + \infty$ for $d < D$ and $\mcm_d^ * = 0$ for $d > D$. \\

\tab The {\em upper} (resp. {\em lower}) {\em Minkowski content} of $\mcl$ is defined by $\mcm^* := \mcm_D^*$ (resp., $\mcm_* := \mcm_{*,D}$). We always have $0 \leq \mcm_* \leq \mcm^* \leq \infty$. If $0 < \mcm_* (\leq) \mcm^* < \infty$, then $\mcl$ is said to be {\em Minkowski nondegenerate.} If, in addition, $\mcm_* = \mcm^*$ (i.e., if the upper limit in Equation \eqref{2.4} is a true limit in $(0, + \infty)$, with $d:=D$), then we denote by $\mcm$ this common value, called the {\em Minkowsi content} of $\mcl$, and the fractal string $\mcl$ (or its boundary $\partial \Omega$) is said to be {\em Minkowski measurable.} So that $\mcm = \lim_{\varepsilon \rightarrow 0^+} V(\varepsilon)/\varepsilon^{1-D}$ and $0 < \mcm < \infty$.\\

\tab We close this section by stating a theorem that shows the intimate connections between the geometric zeta function  $\zeta_\mcl$ of a fractal string $\mcl$ (introduced in \S1 above, just after Equation \eqref{1.1}), and the Minkowski dimension of $\mcl$. (Recall that $\zeta_\mcl (s) = \sum_{j=1}^\infty \ell_j^s$, for all $s \in \mbc$ with $\Res$ sufficiently large.) This result was first observed by the author in [Lap2, Lap3], using earlier work of Besicovich and Taylor in \cite{BesTay}, and has since then been given several direct proofs in [Lap-vFr3, Theorem 1.10] and in [Lap-vFr3, Theorem 13.111]; see also \cite{LapLu-vFr2}.

\begin{theorem}[\cite{Lap2, Lap3}]\label{T2.1}
Let $\mcl$ be a fractal string. Then, the abscissa of convergence of $\zeta_\mcl$ coincides with the Minkowski dimension of $\mcl: \sigma = D$.
\end{theorem}

\tab Recall that the {\em abscissa of convergence} of $\zeta_\mcl$ is given by 

\begin{equation}\label{2.6}
\sigma := \inf \bigg\{ \rho \in \mathbb{R}: \sum_{j=1}^\infty \ell_j^\rho < \infty \bigg\}.
\end{equation} 
It follows from this definition and from known results about Dirichlet series with positive coefficients (see, e.g., [Ser, \S VI.2.2 and \S VI.2.3]) that $\zeta_\mcl = \zeta_\mcl (s)$ is holomorphic for $\Res > D$ and that the open right half-plane $\{\Res > D\}$ is the largest right half-plane (of the form $\{\Res > \alpha \},$ for some $\alpha \in \mathbb{R} \cup \{\pm \infty \}$) to which $\zeta_\mcl$ can be holomorphically continued as well as the largest such half-plane in which the Dirichlet series $\sum_{j=1}^\infty \ell_j^s$ is absolutely convergent (and hence, 
convergent). Therefore, $\zeta_\mcl$ does not have any pole in $\{\Res > D \}.$ (We refer, e.g., to [Ser, \S V1.2 and \S V1.3] or to \cite{HardWr} for an introduction to the theory of Dirichlet series.)\\

\tab Furthermore, note that $s =D$ is always a singularity of $\zeta_\mcl$ (i.e., $\zeta_\mcl (s) \rightarrow + \infty$ as $s \rightarrow D^+, \ s \in \mathbb{R}$) and therefore, if $\zeta_\mcl$ can be meromorphically continued to an open (connected) neighborhood of $D$, then $D$ is a pole of $\zeta_\mcl$. Theorem \ref{T2.1} above (according to which $D = \sigma$), along with this last observation, is one of the original justifications for calling the poles of a meromorphic continuation of $\zeta_\mcl$ (to an open connected neighborhood of $\{\Res > D \}$) the (visible) complex dimensions of the fractal string $\mcl$. (See [Lap-vFr1--3] and, for a brief introduction, see \S 7.2 below.) \\

\tab Here and thereafter, $\{\Res > \alpha \}$ stands for $\{s \in \mathbb{C}: \Res > \alpha \}$. By convention, for $\alpha = -\infty$ or $+ \infty$, respectively, it coincides with $\mathbb{C}$ or the empty set $\emptyset$. Similarly, if $\alpha \in \mathbb{R}, \{\Res = \alpha \}$ denotes the vertical line $\{s \in \mathbb{C} : \Res = \alpha \}.$

\section{Characterization of Minkowski Measurability and Nondegeneracy}

\tab We recall here some of the joint results of the author and Carl Pomerance obtained in [LapPom1,2]. Further results from that work will be discussed in \S5.

\begin{theorem}[Characterization of Minkowski measurability, \cite{LapPom2}]\label{T3.1}
Let $\mcl = (\ell_j)_{j=1}^\infty$ be a fractal string of Minkowski dimension $D \in (0,1)$. Then, $\mcl$ is Minkowski measurable if and only if 
\begin{equation}\label{3.1}
\ell_j \sim Lj^{-1/D} \textnormal{ as } j \rightarrow \infty,
\end{equation}
for some constant $L \in (0, + \infty)$. 
In that case, the Minkowski content of $\mcl$ is given by 
\begin{equation}\label{3.2}
\mcm = \frac{2^{1-D}}{1-D} L^D.
\end{equation}
\end{theorem}

\tab Equation \eqref{3.1} precisely means that the limit of $\ell_j \cdot j^{1/D}$ exists in $(0, +\infty)$ and is equal to $L$. Furthermore, note that \eqref{3.1} is equivalent 
to 

\begin{equation}\label{3.3}
N_\mcl (x) \sim M x^D \textnormal{ as } x \rightarrow + \infty,
\end{equation} 
with $M := L^D$ and where $N_\mcl$, the {\em geometric counting function} of $\mcl$, is defined for all $x > 0$ by

\begin{equation}\label{3.4}
N_\mcl (x) = \# (\{j \geq 1 : \ell_j^{-1} \leq x \}).
\end{equation}
Here and thereafter, $\# B$ denotes the number of elements of a finite set $B$. Moreover, Equation \eqref{3.3} precisely means that $x^{-D} \ N_\mcl (x) \rightarrow M$ as $x \rightarrow +\infty$, for some $M \in (0, + \infty)$.\\

\tab Although we will not use this result explicitly, it is helpful to also give the counterpart of Theorem \ref{T3.1} for Minkowski nondegeneracy. Let $\alpha_*$ (resp., $\alpha^*$) be the lower (resp., upper) limit of $\ell_j \cdot j^{1/D}$ as $j \rightarrow \infty$; so that we always have $0 \leq \alpha_* \leq \alpha^* \leq \infty.$

\begin{theorem}[Characterization of Minkowski nondegeneracy, \cite{LapPom2}]\label{T3.2}
Let $\mcl$ be a fractal string of dimension $D \in (0,1)$. Then, $\mcl$ is Minkowski nondegenerate $($i.e., $0 < \mcm_* (\leq) \mcm^* < \infty)$ if and only 
if

\begin{equation}\label{3.5}
0 < \alpha_* (\leq) \alpha^* < \infty.
\end{equation}
\end{theorem}

\tab Concretely, Equation \eqref{3.5} means that there exist positive constants $c, \ C \geq 1$ such that $c^{-1} j^{-1/D} \leq \ell_j \leq c j^{-1/D}$ for all $j \geq 1$ or, equivalently, $C^{-1} x^D \leq N_\mcl (x) \leq C x^D$ for all $x > 0.$ \\

\tab Let $N_\nu = N_{\nu, \mcl}$ denote the {\em spectral} (i.e., {\em frequency}) {\em counting function} of $\mcl$. Hence, $N_\nu (x) := \# (\{f \in \sigma (\mcl): f \leq x \})$ for all $x >0$, where $\sigma (\mcl) = \{n \cdot \ell_j^{-1} : n \geq 1, j \geq 1 \}$ denotes the ({\em frequency}) {\em spectrum} of $\mcl$. (In essence, the {\em frequencies} of $\mcl$ are, up to a multiplicative normalizing positive constant, equal to the square roots of the eigenvalues of $\mcl$; that is, of the eigenvalues of the Dirichlet Laplacian $- \Delta = -d^2/dy^2$ on any geometric representation $\Omega$ of $\mcl$ by a bounded open subset of $\mathbb{R}$.) It is also shown in \cite{LapPom2} that Equation \eqref{3.5} (and hence also, the Minkowski nondegeneracy of $\mcl$, according to Theorem \ref{T3.2}) is equivalent 
to

\begin{equation}\label{3.6}
0 < \delta_* (\leq) \delta^* < \infty,
\end{equation}
where $\delta_*$ (resp., $\delta^*$) denotes the lower (resp., upper) limit of $\varphi_\nu (x)/x^D$ as $x \rightarrow + \infty$, and $\varphi_\nu (x)$ is the ``asymptotic second term'' for $N_\nu (x)$; namely, $\varphi_\nu (x) := W (x) - N_\nu(x)$, with $W(x):= |\Omega|x$ being the Weyl (or leading) term to be discussed in \S 4.1, \S5 and \S6 below. (In general, one always has $\varphi_\nu (x) \geq 0$ for all $x>0$ and thus $0 \leq \delta_* \leq \delta^* \leq \infty$.)\\

\tab Equation \eqref{3.6} means that there exist $c_1 \geq 1$ such that $c_1^{-1} x^D \leq \varphi_\nu (x) \leq c_1 x^D$ for all $x >0$; in other words, the error estimates of \cite{Lap1}, to be discussed further on in Theorem \ref{T4.1} of \S 4.1, are sharp in this case. More specifically, when $N=1$ and $D \in (0,1)$, each of the equivalent conditions \eqref{3.5} and \eqref{3.6} characterizes  the sharpness of the remainder estimates of \cite{Lap1}, recalled in Equation \eqref{4.3} (see Theorem \ref{T4.1}($i$)) below). \\

\tab It is natural to wonder whether the counterpart of Theorem \ref{T3.2} and of its complement for $N_\nu$ holds for one-sided (rather than two-sided estimates). The answer may be somewhat surprising to the reader. In short, it is positive for the upper estimates but negative for the lower estimates. More specifically, still in \cite{LapPom2}, it is shown 
that

\begin{align}\label{3.7}
&\mcm^* < \infty \Leftrightarrow \ell_j = \ O(j^{-1/D}) \text{ as } j \rightarrow \infty \text{ (i.e., } \alpha^* < \infty) \Leftrightarrow N_\mcl (x) = O(x^D) \\
\notag &\text{ as } x \rightarrow +\infty \Leftrightarrow \varphi_{\nu} (x) = O(x^D) \text{ as } x \rightarrow +\infty, \text{ (i.e., } \delta^* < \infty),
\end{align}
and, more generally, that the exact counterpart of \eqref{3.7} holds for any $d >0$, provided $D$ and $\mcm^* = \mcm_D^*$ are replaced by $d$ and $\mcm_d^*$, respectively. This provides, in particular, a converse (in the present one-dimensional case) to the error estimate obtained in \cite{Lap1} for the asymptotic second term of $N_\nu (x)$ to be discussed next; see part ($i$) of Theorem \ref{T4.1} below, specialized to the case where $N=1$ and $D \in (0,1)$.\\

\tab Moreover, it is shown in \cite{LapPom2} by means of an explicit counterexample that in the analog of \eqref{3.7} for $\mcm_*,$ the implications in one direction holds, but not in the other direction. For example, it is {\em not} true, in general, that $\mcm_* > 0 \Leftrightarrow \alpha_* > 0 \Leftrightarrow \delta_* > 0$. \\

\tab Here, in Equation \eqref{3.7}, as well as in the sequel, given $f: [0, +\infty) \rightarrow \mathbb{R}$ and $g: [0, + \infty) \rightarrow [0, +\infty), $ one writes that $f(x) = O(g(x))$ to mean that there exists a positive constant $c_2$ such that $|f(x)| \leq c_2 g(x),$ for all $x$ sufficiently large. (For the type of functions or sequences we will work with, we may assume that this inequality holds for all $x > 0.$) We use the same classic Landau notation for sequences instead of for functions of a continuous variable.\\

\tab In closing this section, we note that under mild assumptions on $\mcl$ (about the growth of a suitable meromorphic continuation of its geometric zeta function $\zeta_\mcl$), it has since then been shown that within the theory of complex dimensions developed in [Lap-vFr1--3], the characterization of Minkowski measurability obtained in \cite{LapPom2} (Theorem \ref{T3.1} above) can be supplemented as follows (see [Lap-vFr3, Theorem 8.15]), under appropriate hypotheses.

\begin{theorem}[Characterization of Minkowski measurability revisited, \cite{LapPom2, Lap-vFr3}]\label{T3.3}
Let $\mcl$ be a fractal string of Minkowski dimension $D$. Then, under suitable  conditions on $\zeta_\mcl$ $($specified in 
$[$Lap-vFr3, Section 8.3$])$, the following statements are equivalent$:$

\begin{itemize}
\item[$($i$)$] \quad $\mcl$ is Minkowski measurable.
\item[$($ii$)$] \quad Condition \eqref{3.1}, or equivalently \eqref{3.3}, holds.
\item[$($iii$)$] \quad $D$ is the only complex dimension of $\mcl$ with real part $D$, and it is simple.
\end{itemize}
Moreover, if any of these conditions is satisfied, then the Minkowski content of $\mcl$ is given by
\begin{equation}\label{3.8}
\mcm = 2^{1-D} \frac{M}{1-D} = 2^{1-D} \ \frac{res (\zeta_\mcl, D)}{D(1-D)},
\end{equation}
where $res (\zeta_\mcl, D)$ denotes the residue of $\zeta_\mcl (s)$ at $s=D, \ M := L^D$ and $L$ and $M$ are given as in \eqref{3.1} and \eqref{3.3}, respectively.
\end{theorem}

\tab Recall that the complex dimensions of $\mcl$ are the poles of a (necessarily unique) meromorphic continuation of $\zeta_\mcl$ (to a connected open neighborhood of $\{ \Res > D\}$). According to a result of [Lap2,3] (see Theorem \ref{T2.1} above), there are no complex dimensions with real parts $> D$; see the discussion following Equation \eqref{2.6} above.

\begin{rem}\label{R4.5}
When $\mcl$ is a self-similar string $($in the sense of $[$Lap-vFr3, Chapter 2$]$ i.e., the boundary of $\mcl$ is a self-similar set$)$, it is shown in $[$Lap-vFr3, Theorems 2.16, 8.23 and 8.36$]$ that no hypothesis on $\zeta_\mcl$ is needed in the counterpart of Theorem \ref{T3.3}, and that either of the equivalent statements $($i$)$, $($ii$)$ or $($iii$)$ of Theorem \ref{T3.3} is true if and only if $\mcl$ is nonlattice; i.e., iff the subgroup of $\mbr$ generated by the logarithms of its distinct scaling ratios is not of the form $\rho \mathbb{Z}$, for some $\rho >0$. If $\mcl$ is lattice, then it is Minkowski nondegenerate but is not Minkowski measurable.
\end{rem} 

\tab Finally, we mention that the original proof of Theorem \ref{T3.1} given in [LapPom1,2] was analytical and combinatorial in nature. Several parts of the proof have since been established in a different manner by Falconer in \cite{Fal2}, using some techniques from the theory of dynamical systems, and more recently (and perhaps most concisely), by Rataj and Winter in \cite{RatWi2} using techniques from geometric measure theory (in particular, from \cite{Sta, RatWi1} and the relevant references therein).

\begin{rem}\label{R3.4} 
We also note that the notion of Minkowski dimension was introduced $($for noninteger values of the dimension$)$ by Bouligand in the late 1920s in $\cite{Bou}$. The notion of $($normalized$)$ Minkowski content was introduced by Feder in $\cite{Fed}$, while that of Minkowski measurability was apparently first used by Stach\' o in $\cite{Sta}$.\\
\indent \indent The Minkowski dimension is also often called ``box dimension'' $($see, e.g., $\cite{Fal1, Lap-vFr3, Man, Mat, Tri, LapRo2}$ or the applied literature on fractal \\dimensions$)$, entropy dimension, or capacity dimension, for example. Contrary to the Hausdorff dimension, it is not $\sigma$-stable $($that is, it is usually not true that $D (\cup_{k=1}^\infty A_k) = \sup_{k \geq 1} D (A_k))$, where $D(A)$ denotes the upper Minkowski dimension of $A;$ see, e.g., $\cite{Fal1, Mat, 
Tri}$.\footnote{A simple counterexample is provided by $A := \{1/k: k \geq 1 \}$ and $A_k := \{\frac{1}{k} \}$ for each $k \geq 1$, viewed as subsets of $\mbr$; then, $D(A) = 1/2,$ whereas $D(A_k) =0$ for all $k \geq 1$ and hence, $\sup_{k \geq 1} D(A_k) = 0.$} 
Furthermore, unlike the Hausdorff measure $($which is a true positive Borel measure, in the usual mathematical sense of the term, $\cite{Coh, Fal1, Foll, Mat, LapRo2})$, the Minkowski content $($when it exists$)$ or more generally, the upper Minkowski content, is not a measure $($it is only finitely sub-additive$)$. In fact, as is pointed out in $\cite{Lap1}$ $($see also $[$Lap2--3$])$, and somewhat paradoxically, this is precisely because it does not have all of these desirable mathematical properties that the Minkowski dimension is important in the study of aspects of harmonic analysis as well as of spectral and fractal geometry, including the study of the vibrations and spectra of fractal drums or ``drums with fractal boundary'' $([$Lap1--3$]$, see also \S4 below$)$ and, in particular, of fractal strings $($i.e., one-dimensional drums with fractal boundary$)$. See, e.g., $[$Lap1, Examples 5.1 and 5.1'$].$
\end{rem} 

\section{The Weyl--Berry Conjecture for Fractal Drums}

In this section, we first briefly recall Weyl's classic formula for the leading spectral asymptotics of ordinary (smooth or piecewise smooth) drums, as well as corresponding extensions and sharp remainder estimates (from \cite{Lap1}), valid for general fractal drums and providing a partial resolution of the Weyl--Berry conjecture ([Berr1--2]) in any dimension; see \S 4.1. (In \S5, we will specialize the situation to the one-dimensional case, which is the main focus of this paper.) In the latter part of this section, we will comment on various aspects of Weyl's formula for fractal drums, both in the case of drums with fractal boundary (which is of most interest here) and in the related case of drums with fractal membrane (corresponding to Laplacians and Dirac operators on fractals themselves); see \S 4.2. We will also conclude this section by briefly mentioning some of the physical and technological applications of these mathematical results about fractal drums and of the original (or modified) Weyl--Berry conjecture.

\subsection{Weyl's asymptotic formula with sharp error term for fractal drums} 

Hermann Weyl's classic asymptotic formula ([We1,2]; see also, e.g., \cite{CouHil, ReSi3}) for the frequency (or spectral) counting function $N_\nu = N_\nu (x)$ of an ordinary ($N$-dimensional) drum can be stated as follows:

\begin{equation}\label{4.1}
N_\nu(x) = \mathcal{C}_N |\Omega|_N x^N + o(x^N)
\end{equation}
as $x \rightarrow +\infty$. Here, $|\Omega| := |\Omega|_N$ denotes the volume of the bounded open set $\Omega \subset \mathbb{R}^N$ (i.e., the $N$-dimensional Lebesgue measure of $\Omega$). Furthermore, given $x>0, N_\nu (x)$ is the number of (suitably normalized) frequencies $f$ of the drum not exceeding $x$, and $\mathcal{C}_N$ is an explicitly known positive constant which can be expressed in terms of the volume of the unit ball of $\mathbb{R}^N$ (and therefore, in terms of appropriate values of the gamma function $\Gamma = \Gamma(s)$). In the sequel, we denote the {\em leading term} in \eqref{4.1} by 

\begin{equation}\label{4.2}
W(x) := \mcc_n |\Omega|_N x^N
\end{equation} 
and call it the {\em Weyl term.} In addition, mathematically, the spectrum of the ($N$-dimensional) ``drum'' is interpreted as the spectrum of the Dirichlet Laplacian $-\Delta$ on a given nonempty bounded open set $\Omega \subset \mathbb{R}^N (N \geq 1)$, with boundary $\partial \Omega$. Furthermore, the Dirichlet boundary conditions are interpreted variationally (or in the distributional sense); see, e.g., \cite{LioMag, Bre} or [Lap1, \S2].\\

\tab Since the Laplacian is a second order (self-adjoint, positive) linear operator, the aforementioned (normalized) frequencies are (up to a multiplicative constant) equal to the square roots of the eigenvalues. In the present situation, the spectrum is discrete and hence, we can order these frequencies in nondecreasing order (and according to their multiplicities) as follows:

\begin{equation}\label{17.5}
0 < f_1 \leq f_2 \leq \cdots \leq f_n \leq \cdots, \text{ with } f_n \rightarrow + \infty \text{ as } n \rightarrow \infty.
\end{equation}

\begin{rem}\label{R6.5}
Physically, $W(x)$ can be interpreted as a volume in phase space. More specifically, $W(x)$ is proportional to \[ |\{(x, \xi) \in \Omega \times \mathbb{R}^N : |\xi|^2 \leq x^2 \}|_{2N} = |B(0,1)|_N \ |\Omega|_N \ x^N,\] where $\mbr^N \times \mbr^N \approx \mbr^{2N}$ is the ``phase space'' $($the space of positions and velocities or equivalently, momenta, of the classical particle$)$ and for $\rho >0, B(0, \rho)$ denotes the ball of center the origin and radius $\rho$ in $\mbr^N$.
\end{rem}

\tab In \cite{Berr1, Berr2}, extending to the fractal case a classic conjecture of the mathematician Hermann Weyl [We1,2] in the ``regular'' (or ``smooth'' 
case),\footnote{See, e.g., [Mel1--2, Ivr1--3] and [H\" o1--3, Ph, See1--3] (along with the relevant references therein and in [Lap-vFr3, \S 12.5 and Appendix B]) for results (in the smooth case) concerning the Weyl conjecture about the asymptotic second term of the spectral counting function $N_\nu (x)$.} 
the physicist Michael Berry has conjectured that \eqref{4.1} should be completed to obtain an asymptotic second term for the spectral counting function $N_\nu (x)$, of the form $-C_{N,H} \mathcal{H}_H(\partial \Omega) x^H =: S(x)$, where $H$ is the Hausdorff dimension of the boundary $\partial \Omega$ ($H \in [N-1,N]), \mathcal{H}_H (\partial \Omega)$ is the $H$-dimensional Hausdorff measure of $\partial \Omega$ (a well-known fractal generalization to noninteger dimensions of the notion of ``volume''; see, e.g., \cite{Fed, Fal1, Mat, Tri, LapRo2}), and $C_{N,H}$ is a positive constant independent of $\Omega$ and depending only on $N$ and $H$ (as well as expressed in terms of the gamma function, by analogy with the known results in integer dimensions for simple examples such as $N$-dimensional cubes).\\

\tab Unfortunately, it turns out that Berry's conjecture (called the Weyl--Berry conjecture in \cite{Lap1} and in the literature since then), although very stimulating, is not correct, as was first noted by Brossard and Carmona by means of an explicit counterexample in \cite{BroCar} and then, explained from a mathematical point of view (and illustrated by a family of even simpler counterexamples) in \cite{Lap1}. (See, in particular, [Lap1, Examples 5.1 and 5.1'].) Furthermore, $H$ should be replaced by $D$, the (inner) Minkowski dimension of $\partial \Omega$, and $\mathcal{H}_H (\partial \Omega)$ might reasonably be replaced by $\mcm_D (\partial \Omega),$ the (inner) Minkowski content of $\partial \Omega$. Finally, as was shown in [Lap2,3] and [LapPom2,3], even the expected constant $C_{N,H}$ (when it exists) does not simply take the form of $C_{N,D}$ (where $D$ is the Minkowski dimension of $\partial \Omega$) but whatever replaces the factor of proportionality in the counterpart of $S(x)$ should merely be expressed in terms of the residue at $s =D$ of the meromorphic continuation of the corresponding spectral zeta function $\zeta_\nu (s) = \sum_{n=1}^\infty f_n^{-s}$, where $(f_n)_{n=1}^\infty$ is the sequence of frequencies of the drum, as given in Equation \eqref{17.5} above. In fact, in \S5, we will see that when $N=1$ (the special case of fractal strings instead of higher-dimensional fractal drums) and when $\partial \Omega$ (or, equivalently, the fractal string) is Minkowski measurable (with Minkowski content denoted by $\mcm$), then $C_{1,D}$ is proportional to the positive number $-\zeta (D)$, where $\zeta = \zeta (s)$ is the classic Riemann zeta function and $D \in (0,1)$. (See Theorem \ref{T5.3}  below, where in light of \eqref{5.3}, $C_{1,D}$ is of the form $c_D \mcm$, with $c_D$ given by Equation \eqref{5.4}.)\\

\tab On the positive side, the following partial resolution of the Weyl--Berry conjecture was obtained by the author in \cite{Lap1} (recall that the Weyl term $W(x)$ is given by \eqref{4.2} above and is therefore proportional to $x^N$): 

\begin{theorem}[Sharp error estimates, \cite{Lap1}]\label{T4.1}
Let $\Omega$ be any $($nonempty$)$ bounded open subset of $\mathbb{R}^N$. Recall that we always have $D \in [N-1, N]$, where $D = D(\partial \Omega)$ is the $($inner$)$ Minkowski dimension of $\partial \Omega;$ see comment $($b$)$ in \S 4.2 just below.\\

\tab $($i$)$ Then, in the ``fractal case'' where $D \in (N-1, N],$ we have
\begin{equation}\label{4.3}
N_\nu (x) = W(x) + O(x^D) \textnormal{ as } x \rightarrow + \infty,
\end{equation}
provided $\mcm_D^* (\partial \Omega) < \infty$, where the $($pointwise$)$ remainder estimate $O(x^D)$ is sharp for every $D\in (N-1, 
N)$. Furthermore, if $\mcm_D^* (\partial \Omega) = + \infty$, then $D$ should be replaced by $D + \varepsilon$, for any arbitrarily small $\varepsilon > 
0.$\\

\tab $($ii$)$ In the ``nonfractal case'' where $D = N-1$ $($which is the case, for example, if the boundary $\partial \Omega$ is piecewise smooth or, more generally, locally Lipschitz$)$, then exactly the same result as in part $($i$)$ holds, except for the fact that the error term now takes the form $O(x^D \log x)$ as $x \rightarrow + \infty$ $($with $D := N-1)$. 
\end{theorem}

\tab When $\mcm_D^* (\partial \Omega) = + \infty$, either in case ($i$) or ($ii$) of Theorem \ref{T4.1}, one should try to use the later extension of this theorem obtained in \cite{HeLap} and expressed in terms of generalized (upper) Minkowski contents, relative to suitable gauge functions; see comment ($g$) in \S 4.2 just below.\\

\tab Note that in the most fractal case when $D=N$, the remainder estimate \eqref{4.3} still holds but is clearly uninteresting because then, the error term is of the same order as the leading term $W(x)$ given by Equation \eqref{4.2}. 

\subsection{Further comments, extensions and applications}

We next comment on various aspects and extensions (in related contexts) of Theorem \ref{T4.1} and of the Weyl--Berry conjecture. Just as in \S 4.1, our discussion is not meant to be exhaustive in any way but simply aims at providing various pointers and references where the interested reader can find a lot more detailed information.\\

\tab ($a$) The nonfractal case where $D := N-1$ (see part ($ii$) of Theorem \ref{T4.1} just above) was essentially already obtained by G. M\' etivier in \cite{Met3} (see also [M\' et1,2]), in a slightly less precise form and using a different terminology (not explicitly involving the notions of Minkowski dimension and content).\\

\tab ($b$) In Theorem \ref{T4.1}, the (inner) Minkowski dimension $D$ and (inner) upper Minkowski content $\mcm_D^* := \mcm_D^* (\partial \Omega)$ are defined exactly as in the one-dimensional case  where $N=1$ (see Equations \eqref{2.5} and \eqref{2.4}, respectively), except that on the right-hand side of \eqref{2.4} (with $d:=D$), $V(\varepsilon)/\varepsilon^{1-D}$ should be replaced by $V(\varepsilon)/\varepsilon^{N-D}$. Since $\Omega \subset \mbr^N$ is bounded, it has finite volume and therefore it is easy to check that $D \leq N$. Furthermore, since $\partial \Omega$ is the boundary of a (nonempty) bounded open set, its topological dimension is equal to $N-1$ and hence, $D \geq N-1$. Consequently, as was observed in \cite{Lap1}, we always have $N -1 \leq D \leq 
N$. (For the special case of fractal strings, we have $N=1$ and we thus recover the fact that $0 \leq D \leq 1$; see the statement following Equation \eqref{2.5} in \S2.) \\

\tab In \cite{Lap1}, the case where $D= N-1$ is called the {\em least} or {\em nonfractal case}, the case where $D=N$ is called the {\em most fractal case}, while the case where $D = N - \frac{1}{2}$ (i.e., the {\em codimension} $N-D = \{D \}$ is equal to $1/2$, where $\{D \} \in [0,1)$ is the fractional part of $D$), is referred to as the {\em midfractal case.} It turns out that each of these cases (where the codimension $N-D$ takes the value $0,1$ or $1/2$, 
respectively)\footnote{Strictly speaking, when $D = N-1, N-D =1$ is not equal to $\{ 1\}$.} 
plays an important role in the proof of Theorem \ref{T4.1} (and its generalizations) given in \cite{Lap1} (as well as in related spectral or geometric results obtained in [Lap1, Examples 5.1 and 5.1', along with Appendix C]).\\

\tab ($c$) To show that the remainder estimates are sharp (in case ($i$) of Theorem \ref{T4.1} where $D \in (N-1, N)$ and $\mcm_D^* (\partial \Omega) < \infty$), a simple one-family of examples is constructed in [Lap1, Example 5.1 $(N=1)$ and Example 5.1' $(N \geq 1)$]. When $N=1$, it is of the form $\Omega_a := \cup_{j=1}^\infty ((j+1)^{-a}, j^{-a}),$ where $a > 0$ is arbitrary, so that $\partial \Omega_a = \{0\} \cup \{j^{-a}: j \geq 1 \}. $ It follows that for every $a > 0$, $H=0$ (since $\partial \Omega_a$ is countable), whereas $D = (a +1)^{-1}$, $\partial \Omega_a$ is Minkowski measurable (as defined in \S2 above and shown in \cite{Lap1}) with Minkowski content $\mcm := \mcm_D (\partial \Omega)$ equal to $2^{1-D} a^D/(1-D)$; see [Lap1, Appendix C]. This was the first explicit example of a fractal string and served as a motivation for the formulation of Theorem \ref{T3.1} above (the characterization of Minkowski measurability for fractal strings, obtained in \cite{LapPom2}) and the modifed Weyl--Berry conjecture, stated in \cite{Lap1} and to be discussed in \S5 below (see Conjecture \ref{C5.1}). Note that the midfractal case where $D= 1/2$ corresponds to $a=1$, while the most ($D=1$) and least ($D=0$) fractal cases correspond, respectively, to the limits $a \rightarrow 0^+$ and $a \rightarrow + \infty$. Furthermore, the symmetry $a \leftrightarrow 1/a$ exchanges $D$ and $1-D$. \\

\tab Finally, when $N \geq 2$, exactly the same statements as above are true for the $N$-dimensional analog (a fractal comb) of $\Omega_a$, defined by $\Omega_{a,N} := \Omega_a \times [0,1]^{N-1}$. We then have that $H = N-1, D= N-1 + (a +1)^{-1},$ $\partial \Omega_{a, N}$ is Minkowski measurable and $\mcm_D(\partial \Omega_{\alpha, N})$ has the same value as above. Furthermore, the remainder estimate \eqref{4.3} of part ($i$) of Theorem \ref{T4.1} is still sharp for every $a>0$ (and hence, for every $D \in (N-1, N)$); see [Lap1, Example 5.1']. Actually, it follows from the later results (from \cite{LapPom2}) recalled in \S5 below that for every $a>0$,

\begin{equation}\label{4.4}
N_\nu (x) = W(x) - c_{N,D} \ \mcm_D (\partial \Omega) x^D + o(x^D) \text{ as } x \rightarrow + \infty, 
\end{equation}
where $\Omega = \Omega_{a, N}$, $W(x)$ is given by \eqref{4.2}, and $c_{N,D}$ is explicitly known.\\

\tab ($d$) Under suitable hypotheses, Theorem \ref{T4.1} has an exact counterpart for the Neumann Laplacian (instead of the Dirichlet Laplacian), with the boundary conditions interpreted variationally (as in \cite{Bre, LioMag} or [Lap1, \S2]) as well as for higher order, positive self-adjoint elliptic operators with (possibly) variable coefficients and with Dirichlet, Neumann or mixed boundary conditions; see \cite{Lap1}, Theorem 2.1 and its corollaries. For the Neumann Laplacian, the error estimate \eqref{4.3} holds, for instance, for the classic Koch snowflake domain and, more generally, for all planar quasidics \cite{Pomm, Maz} (e.g., for the simply connected planar domains bounded by the Julia sets of the quadratic maps $z \mapsto z^2 + c$, where the complex parameter $c$ is sufficiently small).\\

\tab ($e$) In \cite{BroCar}, for the Dirichlet Laplacian and using probabilistic methods, Brossard and Carmona have obtained error estimates (as $t \rightarrow 0^+$) for the partition function $Z_\nu = Z_\nu (t) := Tr (e^{t \Delta})$ (the trace of the heat semigroup), of interest in quantum statistical mechanics, probability theory, harmonic analysis and spectral theory. These estimates are also expressed in terms of the (inner) Minkowski dimension $D$. We  should note, however, that even though part ($i$) of Theorem \ref{T4.1} implies those error estimates, the converse is not true, in general. Indeed, typically, beyond the leading term (for which a classic Tauberian theorem can be used in order to show the equivalence of the corresponding results for $N_\nu (x)$ and $Z_\nu (t)$ (see, e.g., \cite{Kac}, \cite{Sim}), pointwise asymptotic results for $N_\nu (x)$ are considerably more difficult to obtain than for $Z_\nu (t)$. \\

\tab ($f$) The Weyl--Berry conjecture for drums with fractal boundary (and its modifications in \cite{Lap1}) has since been studied  in a number of different contexts, both analytically and probabilistically. We refer to [Lap-vFr3, \S 12.5] for a number of references on the subject, including [BroCar, Lap1--3, LapPom1--3, LapMai1--2, Ger, GerSchm, FlVas, HeLap, MolVai, vBGilk, Lap-vFr1--3, HamLap, LapRa\u Zu1, LapRa\u Zu7]. We also refer to \cite{Gilk}, [H\" o1--3], [Lap1--3] and [Lap-vFr3, Appendix B] for related references (including \cite{Gilk} in the case of partition functions) on the spectral asymptotics of smooth (as opposed to fractal) drums. A general introduction to Weyl's aymptotic formula and its analytic or probabilistic proof (for smooth or piecewise smooth boundaries, for example) can be found in \cite{CouHil, ReSi3, Gilk, Kac, Sim}. \\

\tab Finally, we mention that the original Weyl--Berry conjecture was also formulated for ``drums with fractal membrane'' (as opposed to ``drum with fractal boundary''). An appropriate modification of the conjecture was established in that setting by Jun Kigami and the author in \cite{KiLap1} (building on a conjecture of \cite{Lap3}) for a large class of self-similar fractal drums corresponding to Laplacians {\em on} fractals (the so-called ``finitely ramified'' or p.c.f. self-similar sets), such as the Sierpinski gasket, the pentagasket and certain fractal trees. In particular, in \cite{KiLap1} is established a suitable analog of Weyl's classic asymptotic formula, but now for the leading spectral asymptotics of Laplacians on (self-similar) fractals (see, e.g., \cite{Ki}) rather than on bounded open sets with fractal boundary. The resulting semi-classical formula, along with its geometric interpretation, is further explored and extended from the point of view of nonsmooth geometric analysis and Connes' noncommutative geometry \cite{Con}, in [Lap4--5], \cite{KiLap2} and (in a somewhat different setting and using suitably constructed Dirac operators and intrinsic geodesic metrics on the fractals under consideration), in \cite{ChrIvLap} and \cite{LapSar}.  \\

\tab Beside \cite{KiLap1} and \cite{Lap3}, see also, e.g., [Lap-vFr3, \S 12.5.2] for many other references on (or related to) this subject, including [Ram, RamTo, Sh, FukSh, Ham1--2, Lap4--5, KiLap2, Sab1--3, Tep1--2, DerGrVo, ChrIvLap, LalLap1--2, LapSar].\\

\tab ($g$) Theorem \ref{T4.1} above (from \cite{Lap1}), along with the results from \cite{LapPom2} stated in \S3 above and \S5 below, as well as a part of the results from \cite{LapMai2} discussed in \S6 below, have been extended by Christina He and the author in the research memoir \cite{HeLap} to the more general situation where the (now {\em generalized}) Minkowski content is no longer defined in terms of a strict power law, but in terms of a large class of ``gauge functions'' involving, for example, expressions of the form $x^D \log x, \ x^D \log \log x, \cdots$ or $x^D/\log \log \log x$, etc. This is of interest in the applications to fractal geometry, spectral geometry, harmonic analysis, probability theory, stochastic processes and mathematical or theoretical physics.\\

\tab We close this section by briefly commenting on the physical relevance of Weyl's asymptotic formula (see \cite{Kac} for an interesting overview), Theorem \ref{T4.1} above (from \cite{Lap1}) and the Weyl--Berry conjecture which partially motivated it. The Weyl--Berry conjecture, along with its various modifications and extensions, has potential or actual physical and engineering applications to condensed matter physics, quantum mechanics, quantum chemistry, quantum chaos, acoustics, diffusions and wave propagation in fractal or random media  (i.e., on or off fractals), geophysics, radar and cell phone technology (fractal antennas), as well as the making of computer microchips. (See, e.g., [Berr1--2, Kac, Ram, RamTo, Sch, Lap1--6, SapGoMar, KiLap1--2, LapPa, LapNeuReGr, ChrIvSar, LalLap1--2, LapSar] and the relevant references therein.) \\

\tab We will next focus on the one-dimensional case (i.e., $N=1$), as in the rest of this paper (with the exception of the present section), and therefore on the connections between this subject and aspects of number theory, particularly the Riemann zeta function and the Riemann hypothesis, in \S5 and \S6, respectively.

\section{Direct Spectral Problems for Fractal Strings and the Riemann Zeta Function in the Critical Interval}

Having discussed in \S4 the Weyl classical asymptotic formula (with error term) and the closely related Weyl--Berry conjecture, we may now specialize the situation to the one-dimensional case (that of fractal strings, corresponding to $N=1$ in \S 4.1) and discuss the key result of \cite{LapPom2} establishing the (one-dimensional) ``modified Weyl--Berry conjecture'' for fractal strings; see Theorem \ref{T5.3} below. This brings to the fore direct connections between the spectra of fractal strings and the Riemann zeta function $\zeta = \zeta(s)$, in the case of the ``critical interval'' $0 < s <1$. (See also Equation \eqref{1.1} above for a related general formula.) This connection will be further explored in \S6 (based on \cite{LapMai2}), in relation with the critical strip $0< \Res < 1$, the Riemann hypothesis and the converse of Theorem \ref{T5.3}; that is, an inverse (rather than a direct) spectral problem for fractal strings; see, in particular, Theorems \ref{T6.1} and \ref{T6.2} (from \cite{LapMai2}) in \S6 below.\\

\tab Let $\Omega$ be an arbitrary (nontrivial) fractal string (i.e., a bounded open subset of $\mathbb{R}$), of Minkowski dimension $D \in (0,1)$ and associated sequence of lengths $\mcl = (\ell_j )_{j=1}^\infty$, written in nonincreasing order (according to multiplicities) and such that $\ell_j \downarrow 0$ as $j \rightarrow \infty$; see \S2. In the sequel, we will refer to such a fractal string as $\Omega$ or $\mcl$, interchangeably. \\

\tab Recall from our discussion in \S1 and \S 4.1 that a one-dimensional fractal drum (i.e., drum with fractal boundary) is nothing but a fractal string. Furthermore, recall from \S2 that the (Minkowski) dimension of a fractal string always satisfies $0 \leq D \leq 1$. In the sequel, we exclude the extreme cases where $D=0$ and $D=1$ (the least or nonfractal case and the most fractal case, respectively) and therefore assume that $D$ belongs to the {\em critical interval} $(0,1): 0<D<1$.\\

\tab Letting $N=1$ in the expression \eqref{4.2} for the Weyl term (i.e., the leading term in Weyl's asymptotic formula \eqref{4.1}) $W= W(x)$, we obtain (given our current normalization for the frequencies)

\begin{equation}\label{5.1}
W(x) = |\Omega|_1 \ x, \text{ for } x >0.
\end{equation} 
So that, according to Weyl's classic formula \eqref{4.1}, we have

\begin{equation}\label{5.2}
N_\nu (x) = W(x) + R(x),
\end{equation}
where $W(x)$ is given by \eqref{5.1} and $R(x) = o(x)$ as $x \rightarrow + \infty$. Furthermore, provided $\mcm^* < \infty$ (i.e., $\mcl$ has finite upper Minkowski content), then according to part ($i$) of Theorem \ref{T4.1} above (from \cite{Lap1}), the error term $R(x)$ in \eqref{5.2} can be estimated as follows: $R(x) = O(x^D)$ as $x \rightarrow + \infty$. Moreover, if $\mcl$ is Minkowski nondegenerate (i.e., $0< \mcm_* < \mcm^* < \infty$), we even have that $R (x)$ is truly of the order of $x^D$ as $x \rightarrow + \infty$ (according to a result of \cite{LapPom2} briefly discussed in \S2 above; and conversely (still by \cite{LapPom2}), if $R(x)$ is exactly of the order of $x^D$ (resp., if $R(x) = O(x^D)$), then $\mcl$ is Minkowski nondegenerate (resp., $\mcm^* < \infty$). (See Theorem \ref{T3.2} and Equation \eqref{3.7}.) \\

\tab The question is now to know when we really have an asymptotic second term, proportional to $x^D$, in Weyl's asymptotic formula \eqref{5.2}. We will sometimes refer to such a term as a ``{\em monotonic second term}''. In \cite{Lap1}, the following conjecture (called the MWB conjecture, for short) was stated.

\begin{conj}[Modified Weyl--Berry conjecture, \cite{Lap1}]\label{C5.1}
If $\mcl$ is a Minkowski measurable fractal string of dimension $D \in (0,1)$, then 
\begin{equation}\label{5.3}
N_\nu (x) = W(x) - c_D \ \mcm x^D + o(x^D), \textnormal{ as } x \rightarrow + \infty,
\end{equation}
where $\mcm$ denotes the Minkowski content of $\mcl$ (i.e., of its boundary $\partial \Omega)$ and $c_D$ is a positive constant depending only on $D$.
\end{conj} 

\begin{rem}\label{R5.2}
$($a$)$ \ In higher dimensions $($i.e., when $N \geq 2$ and $\Omega \subset \mathbb{R}^N$ is a bounded open set$)$, the counterpart of the MWB conjecture is not true, in general$;$ see $\cite{FlVas}$ and, especially, $\cite{LapPom3}$ for various counterexamples. Moreover, one does not know whether it is true for a simply connected domain in the plane $($i.e., when $N=2)$. It is not the object of the present paper to discuss this issue further, although it is of significant interest and is also very intricate.\\

$($b$)$ \ When $N=1$, the Minkowski measurability condition is necessary, in general, for the spectral counting function $N_\nu$ to admit a {\em monotonic} asymptotic second term, proportional to $x^D$, as $x \rightarrow + \infty$. For example, for the Cantor string $\mcl = \mcl_{CS}$ $($defined by $\Omega_{CS} := [0,1] \ \backslash C$, the complement in $[0,1]$ of the classic ternary Cantor set $C)$, $N_\nu$ admits an {\em oscillatory} $($asymptotic$)$ second term. More specifically, for the Cantor string, the asymptotic second term is of the form $-x^D \ G(\log_3 x)$, where $G$ is a $1$-periodic function which is bounded away from zero and infinity$;$ see $[$LapPom2$]$ and, especially, $[$Lap-vFr3, Equation $($16.57$)$ and \S 10.2.1$]$. This fact was first proved in $[$LapPom1,2$]$ $($by a direct computation$)$ and does not contradict the MWB conjecture since $($as was also first proved in $[$LapPom1,2$]),$ the Cantor string is {\em not} Minkowski measurable $($but is Minkowski nondegenerate$)$. Moreover, these issues were investigated in great detail in $[$Lap-vFr1--3$]$, by using the theory of complex dimensions and associated explicit formulas developed in those monographs. $($See, e.g., $\cite{Lap-vFr3}$, Chapter 6, \S 8.4 and Chapter 10.$)$\\
\indent \indent It is shown, for example, in $[$Lap-vFr3, \S 6 and \S 8.4$]$ $($building in part on conjectures and results in $\cite{Lap3})$ that a self-similar string is Minkowski measurable if and only it is nonlattice $($i.e., the logarithms of its distinct scaling ratios are rationally independent$)$. Moreover, it follows from {\em loc. cit.} that a lattice self-similar string $($e.g., the Cantor string$)$ is never Minkowski measurable $($but is Minkowski nondegenerate$)$ and that its spectral counting function $N_\nu (x)$ always has an asymptotic second term which is {\em oscillatory} and of the order of $x^D$ $($in fact, it is of the form $x^D \ G (\log x)$, where $G$ is a nonconstant periodic function on $\mbr$ which is bounded away from zero and infinity$)$.\\
\indent \indent More generally, we will see in Theorem \ref{T6.1} of \S6 $($based on $\cite{LapMai2})$ that if $\zeta = \zeta(s)$ does not have any zeros on the vertical line $\{\Res = D \}$, then the existence of an asymptotic second term proportional to $x^D$ for the spectral counting function $N_\nu (x)$ implies that $\mcl$ is Minkowski measurable $($i.e., the hypothesis of Conjecture \ref{C5.1} is also necessary$)$.
\end{rem}

\tab The following theorem (due to the author and Carl Pomerance in \cite{LapPom2}, and announced in \cite{LapPom1}) gives the precise asymptotic second term of the spectral (or frequency) counting function of a Minkowski measurable fractal string, thereby resolving in the affirmative the MWB conjecture for fractal strings and yielding the specific value of the positive constant $c_D$ appearing in Equation \eqref{5.3} above.

\begin{theorem}[Resolution of the MWB conjecture for fractal strings, \cite{LapPom2}]\label{T5.3}
The modified Weyl--Berry conjecture for fractal strings $($Conjecture \ref{C5.1} above$)$ is true for every $D \in (0,1)$. More specifically, if $\mcl$ is a Minkowski measurable fractal string of dimension $D \in (0,1)$, then its frequency counting function $N_\nu = N_\nu (x)$ admits a monotonic second term, proportional to $x^D$, of the exact same form as in Equation \eqref{5.3}. Furthermore, the positive constant occurring in \eqref{5.3} is given by

\begin{equation}\label{5.4}
c_D = (1-D) \ 2^{-(1-D)} (-\zeta(D)).
\end{equation}
\end{theorem}

\tab Note that $c_D > 0$ because $\zeta = \zeta(s)$ is strictly negative in the critical interval $0<s< 1$ (see, e.g., \cite{Ti}): $-\zeta (D) > 0$ since $D \in (0,1)$. Furthermore, in Equation \eqref{5.4}, the constant $c_D$ is proportional to the positive number $-\zeta (D)$, and hence, to the value at $s=D$ of $\zeta = \zeta(s)$ in the {\em critical interval} $0 < s < 1.$ \\

\tab We next briefly comment on the structure of the proof of Theorem \ref{T5.3} given in \cite{LapPom2}. It relies on two different theorems. Namely, the geometric characterization of Minkowski measurability (Theorem \ref{T3.1} above, from \cite{LapPom2}) along with the following theorem (also from \cite{LapPom2}), which is of an analytic number theoretic nature and whose motivation will be explained after its statement. (In the sequel, given $y \in \mathbb{R}, [y]$ stands for the {\em integer part} of $y$ and $\{y \} := y - [y] \in [0,1)$ denotes the {\em fractional part} of $y$.)

\begin{theorem}[The sound of Minkowski measurable fractal strings, \cite{LapPom2}]\label{T5.4}
Let $\mcl$ be a Minkowski measurable fractal string of dimension $D \in (0,1)$. Equivalently, according to Theorem \ref{T3.1} $($see Equation \eqref{3.1} in \S3$)$, let $\mcl = (\ell_j)_{j=1}^\infty$ be a nonincreasing sequence of positive numbers satisfying $($for some positive constant $L > 0)$

\begin{equation}\label{5.5}
\ell_j \sim Lj^{-1/D} \textnormal{ as } j \rightarrow \infty.
\end{equation} 
Then 

\begin{equation}\label{5.6}
\sum_{j=1}^\infty [\ell_j x] = \left(\sum_{j=1}^\infty \ell_j \right)x + \zeta(D) L^D x^D + o(x^D) \textnormal{ as } x \rightarrow + \infty, 
\end{equation}
where $\zeta$ denotes the classic Riemann zeta 
function. Equivalently, Equation \eqref{5.6} can be stated as follows$:$ 
\[ \sum_{j=1}^\infty \{\ell_j \ x \} \sim -\zeta (D) \ L^D x^D \textnormal{ as } x \rightarrow + \infty. \]
\end{theorem}

\tab We note that in order to state Theorem \ref{T5.4}, one does not need to assume that $\mcl$ is Minkowski measurable and therefore, to rely on Theorem \ref{T3.1}. Instead, one can simply assume that \eqref{5.5} (or, equivalently, \eqref{3.3}) holds for some $D \in (0,1)$. Theorem \ref{T3.1} is then used when deducing Theorem \ref{T5.3} from Theorem \ref{T5.4}. \\

\tab Given a fractal string $\Omega = \cup_{j=1}^\infty I_j$ (as in \S2), its spectral counting function $N_\nu =: N_\nu (\Omega, \cdot)$ satisfies $N_\nu (\Omega, \cdot) = \sum_{j=1}^\infty N_\nu (I_j, \cdot)$, with $N_\nu (I_j, \cdot) = [\ell_j \ x]$ for each $j \geq 1$ because $I_j$ is an interval of length $\ell_j$. It follows that $N_\nu (x) = \sum_{j=1}^\infty [\ell_j \ x],$ which is the left-hand side of \eqref{5.6}. In light of the first expression obtained for the Minkowski content $\mcm$ in Equation \eqref{3.8} (and the line following it), one then easily deduces Theorem \ref{T5.3} (and hence, the MWB conjecture for fractal strings, Conjecture \ref{C5.1}) from Theorem \ref{T5.4}. Note that $|\Omega|_1 = \sum_{j=1}^\infty \ell_j$, so that the leading term of $N_\nu (x) = \sum_{j=1}^\infty [\ell_j \ x]$ in \eqref{5.6} coincides with the Weyl term $W(x)$ (given by \eqref{5.1}), as it should. Furthermore, as was alluded just above, by using the expression of $\mcm$ given by the first equality in \eqref{3.8} and eliminating $L^D$ in the asymptotic second term of \eqref{5.6}, one establishes both the existence and the value of the positive constant $c_D$ (explicitly given by Equation \eqref{5.4} of Theorem \ref{T5.3}).\\

\tab Physically, the aforementioned relation, $N_\nu (\Omega, \cdot) = \sum_{j=1}^\infty N_\nu (I_j, \cdot)$, follows from the fact that the intervals $I_j$ comprising the fractal string $\Omega$ (or, more poetically, the strings of the fractal harp) are vibrating independently of one another. Mathematically, it follows from the variational formulation of the underlying eigenvalue problem; see, e.g., \cite{Lap1} and the relevant references therein (including \cite{LioMag} and \cite{ReSi3}).

\section{Inverse Spectral Problems for Fractal Strings and the Riemann Hypothesis}
It is natural to wonder whether the converse of Theorem \ref{T5.3} (or essentially equivalently, of the MWB conjecture for fractal strings, Conjecture \ref{C5.1}) is true. Roughly speaking, this means that if there are no oscillations of order $D$ in the spectrum of $\mcl$, then there are no oscillations of order $D$ in the geometry of $\mcl$. Rephrased: If $N_\nu (x)$ has an asymptotic second term, proportional to $x^D$, is it true that $\mcl$ is Minkowski measurable?\\

\tab More precisely, given $D \in (0,1)$, the inverse spectral problem (ISP)$_D$ under consideration can be stated as follows:\\

\noindent (ISP)$_D$ {\em If $\mcl$ is a fractal string of dimension $D$ such 
that}
\begin{equation}\label{6.1}
N_\nu (x) = W(x) - \mcc \ x^D + o(x^D) \text{ as } x \rightarrow + 
\infty,
\end{equation}
{\em for some nonzero} ({\em real}) {\em constant 
$\mcc$, 
then is it true that $\mcl$ is Minkowski measurable} ({\em or, equivalently, in light of Theorem \ref{T3.1} and the comment following it, that $N_\mcl (x) \sim M \ x^D$ as $x \rightarrow +\infty$, for some positive constant $M$, where $N_\mcl = N_\mcl (x)$ denotes the geometric counting function of $\mcl$})? \\

\tab The above problem, (ISP)$_D$, is called an {\em inverse spectral problem} since given some spectral information about the fractal string $\mcl$ (namely, the existence for $N_\nu (x)$ of a monotonic asymptotic second term, proportional to $x^D$), one asks whether one can recover some geometric information about $\mcl$ (namely, that $\mcl$ is Minkowski measurable). Similarly, the MWB conjecture from \cite{Lap1} (Conjecture \ref{C5.1} above) and its resolution given in Theorem \ref{T5.3} above (from \cite{LapPom2}) fall naturally within the class of {\em direct spectral problems}.\\

\begin{rem}\label{R12.5}
$($a$)$ In the statement of $($ISP$)_D$, it is not necessary to assume that $\mcl$ is of Minkowski dimension $D$ since a result from $[$LapPom2$]$ $($recalled in Theorem \ref{T3.2} above and the comment following it$)$ shows that \eqref{6.1} implies that $\mcl$ is Minkowski nondegenerate $($i.e., $0< \mcm_* \leq \mcm^* < \infty)$ and hence, has Minkowski dimension $D$.\\

$($b$)$ Originally, in $[$LapMai1--2$]$, the error term $o(x^D)$ in the counterpart of Equation \eqref{6.1}, was assumed to be slightly smaller, namely, $O(x^D/ \log^{1+\delta} x)$ as $x \rightarrow +\infty$, for some $\delta >0$. However, an unpublished work of Titus Hilberdink $[$Hil$]$, using an improved version of the Wiener--Ikehara Tauberian theorem $($see, e.g., $[$Pos$])$ used in $[$LapMai1--2$]$, shows that the more general estimate $o(x^D)$ suffices. $($Actually, in $[$LapMai1--2$]$, the better error estimate is only used for the sufficiency part of the proof of Theorem \ref{T6.1}, that is, the part requiring the use of a Tauberian theorem.$)$\\

$($c$)$ For Dirichlet boundary conditions $($which are assumed here in $($ISP$)_D)$, one must necessarily have $\mcc >0$ in Equation \eqref{6.1} because then, we have that $N_\nu (x) \leq W (x)$ for all $x >0$, which implies that $\mcc \geq 0$. Note that by hypothesis, $\mcc \neq 0$.
\end{rem}

\tab In \cite{LapMai2} (announced in \cite{LapMai1}), the author and H. Maier have shown that this family of inverse spectral problems (ISP)$_D$, for $D$ ranging through the critical interval $(0,1)$, is intimately related to the presence of zeros of $\zeta = \zeta(s)$ in the {\em critical strip} $0 < \Res < 1$ (i.e., to the {\em critical zeros} of $\zeta$), and thereby, to the Riemann hypothesis. More specifically, we have the following two theorems (the second one really being a corollary of the first one).

\begin{theorem}[Characterization of the nonvanishing of $\zeta$ along vertical lines in the critical strip, \cite{LapMai2}]\label{T6.1}
Fix $D \in (0,1)$. Then, the inverse spectral problem $($ISP$)_D$ is true $($that is, has an affirmative answer for every fractal string of dimension $D$ satisfying the assumptions of $($ISP$)_D)$ if and only if the Riemann zeta function $\zeta = \zeta(s)$ does not have any zeros along the vertical line $\{ \Res =D \};$ i.e., if and only the ``$D$-partial Riemann hypothesis'' is true.
\end{theorem}

\begin{theorem}[Spectral reformulation of the Riemann hypothesis, \cite{LapMai2}]\label{T6.2}
The inverse spectral problem $($ISP$)_D$ is true for every value of $D$ in $(0,1)$ other than in the midfractal case when $D=1/2$ $($or equivalently, for every $D \in (0,1/2))$ if and only if the Riemann hypothesis is true.
\end{theorem}

\tab Theorem \ref{T6.2} follows at once from Theorem \ref{T6.1} since the Riemann hypothesis states that $\zeta(s) = 0$ with $0 < \Res < 1$ implies that $\Res = 1/2$ (i.e., that $s$ belongs to the critical line $\{\Res = 1/2 \}$). The fact that in Theorem \ref{T6.2}, $D$ can be equivalently assumed to be in $(0, 1/2), (1/2, 1)$ or in $(0,1) \backslash \{1/2 \}$, follows from the functional equation satisfied by $\zeta = \zeta(s)$, according to 
which

\begin{equation}\label{6.2}
\xi (s) = \xi (1-s), \text{ for all } s \in \mathbb{C},
\end{equation}
and hence, 

\begin{equation}\label{27.5}
\zeta(s) = 0 \Leftrightarrow \zeta (1-s) = 0, \text{ for } 0 < \Res < 1.
\end{equation}
Here, $\xi = \xi (s)$ denotes the {\em completed Riemann zeta function} (or the {\em global zeta function} of $\mathbb{Q}$, the field of rational numbers), defined by $\xi (s) := \pi^{-s/2} \Gamma (s/2) \zeta (s)$. For the standard properties of $\zeta$ and of $\xi$, we refer, e.g., to \cite{Edw, Ing, KarVor, Pat, Ti}.\\

\tab The next result follows immediately from Theorem \ref{T6.1} and the well-known fact according  to which $\zeta$ has a zero (and even infinitely many zeros, according to Hardy's theorem \cite{Edw, Ti}, even though it is not needed here) along the critical line $\{\Res = 1/2 \}$.

\begin{cor}[\cite{LapMai2}]\label{C6.3}
The inverse spectral problem $($ISP$)_D$ is not true in the midfractal case when $D=1/2$.
\end{cor}

\tab The following interpretation of Theorem \ref{T6.2} and Corollary \ref{C6.3} has first been proposed in [Lap2, Lap3] and has since been pursued in a different, but closely related context, in [HerLap1--5] and in \cite{Lap7}, as will be briefly discussed towards the end of \S 7.4 below.

\begin{theorem}[The Riemann hypothesis as a mathematical phase transition, \cite{Lap2, Lap3}]\label{T6.4}
The Riemann hypothesis is true if and only if $($ISP$)_D$, the inverse spectral problem for fractal strings, fails to be true $($i.e., fails to have an affirmative answer$)$ only in the midfractal case when $D=1/2$.
\end{theorem}

\tab In [Lap2,3], the author also wondered (in an open problem) whether the mathematical phase transition conditionally (i.e., under RH, and in fact, if and only if RH is true) occurring at $D=1/2$ could be understood (in a suitable model) as a true physical phase transition, of the type occurring in the theory of critical phenomena in statistical physics or quantum field theory. He also asked whether the (then) intuitive notion of fractal ``complex dimension'', underlying the proof of the key part of Theorem \ref{T6.1} (as will be explained next), could not be understood as a ``complexified space-time dimension'', as in Wilson's theory of phase transitions and critical phenomena \cite{Wil}.\\

\tab We now briefly comment on the proof of Theorem \ref{T6.1}, the central result of \cite{LapMai2}, from which Theorem \ref{T6.2} and Corollary \ref{C6.3} (as well as Theorem \ref{T6.4}) follow, given the known properties of the Riemann zeta function. First, note that Theorem \ref{T6.1} consists of two different theorems (and was stated as such in [LapMai1,2]. The part of Theorem \ref{T6.1} corresponding to a sufficient condition (for (ISP)$_D$ to be true) is established by using the Wiener--Ikehara Tauberian theorem (see, e.g., \cite{Pos} or \cite{LapMai2} for the statement of this theorem). That is, assuming that (for a given $D \in (0,1)$), $\zeta(s) \neq 0$ for all $s \in \mbc$ on the vertical line $\{\Res = D \}$, one uses the aforementioned Tauberian theorem in order to show that the inverse problem (ISP)$_D$ has an affirmative answer (for all fractal strings of dimension $D$). On the other hand, in order to establish the converse, namely, the fact that the condition that $\zeta(s) \neq 0$ for all $s \in \mbc$ on the vertical line $\{\Res =D\}$ is necessary for the inverse spectral problem (ISP)$_D$ to be true (i.e., to have an affirmative answer for all fractal strings of dimension $D$), one uses in [LapMai1,2] in a key (but rigorous) manner the intuition (at the time) of complex dimension and its intimate connections with asymptotic oscillatory behavior (both in the geometry and the spectrum), along with Theorem \ref{T3.1} (from \cite{LapPom2}).\footnote{It can now also be systematically understood in terms of the generalized explicit formulas of [Lap-vFr3, Chapter 5]; see [Lap-vFr3, Chapter 9]. The resulting theorems and assumptions, however, are somewhat different.} \\

\tab More specifically, we reason by contraposition. Fix $D \in (0,1)$ and assume that $\zeta (\omega) = 0$, for some $\omega \in \mathbb{C}$ such that $Re(\omega) =D$. Then, due to the basic symmetry of $\zeta$ (namely, $\zeta(\overline{s}) = \overline{\zeta (s)}$, where the bar indicates that we are taking the complex conjugate of the given complex number), we also have $\zeta (\overline{\omega}) = 0$. Let us write $\omega = D + i \tau$, with $\tau \in \mbr$; so that $\overline{\omega} = D - i \tau$. Without loss of generality, we may assume that $\tau > 0$. (Clearly, $\tau \neq 0$ since $\zeta(D) < 0$ because $0 < D < 1$.)\\

\tab Next, for $x >0$, let 

\begin{equation}\label{6.3}
U(x) := x^D + \beta (x^\omega + x^{\overline{\omega}}) = x^D (1 +2\beta \cos(\tau \log x)),
\end{equation}
for some coefficient $\beta >0$ sufficiently small, and let $V(x):= [U(x)]$, the integer part of $U(x)$. It is easy to check that for all $\beta$ small enough, $U(x) > 0$ and $U$ is (strictly) increasing on $(0, +\infty)$; furthermore, the range of $U$ is all of $(0, +\infty)$. Hence, for such values of $\beta$, given any integer $j \geq 1$, we can uniquely define $\ell_j >0$ such that $U(\ell_j) = j$ (and thus, $V(\ell_j) = j$). In this manner, we define a fractal string $\mcl = (\ell_j)_{j=1}^\infty$ with geometric counting function $N_\mcl$ coinciding with $V: N_\mcl = V$. In light of Equation \eqref{6.3}, $N_\mcl = V = [U]$ has sinusoidal (and hence, nontrivial periodic) oscillations, caused by the ``complex dimensions'' $\omega$ and $\overline{\omega}$. Therefore, $N_\mcl (x)$ cannot be asymptotic to $x^D$; equivalently, we do not have $N_{\mcl} (x) \sim M \ x^D$ as $x \rightarrow + \infty$, for some 
$M >0$. (Note that here, according to \eqref{6.3}, we would have to have $M:=1$.) It then follows from Theorem \ref{T3.1} (the characterization of Minkowski measurability, from \cite{LapPom2}) and the comment following it that $\mcl$ is {\em not} Minkowski measurable. Moreover, and using the fact that $\zeta(\omega) = \zeta (\overline{\omega}) = 0$, via a direct 
computation\footnote{This is now proved in [Lap-vFr1--3] by using the (generalized) explicit formulas from [Lap-vFr1--3] and Equation \eqref{1.1}.}
it is shown in \cite{LapMai2} that Equation \eqref{6.1} holds for the fractal string $\mcl$, for some (explicitly known) positive constant $\mcc$. Consequently, the hypothesis \eqref{6.1} of (ISP)$_D$ is satisfied but its conclusion (namely, the Minkowski measurability of $\mcl$) is not; i.e., the inverse spectral problem (ISP)$_D$ cannot be true for this value of $D \in (0,1)$ because it fails to hold for this particular fractal string $\mcl$. This proves that if (ISP)$_D$ is true for some $D \in (0,1)$, we must have $\zeta (s) \neq 0$ for all $s \in \mathbb{C}$ such that $\Res =D$, as desired.\\

\tab We point out that in the above construction, the geometric oscillations caused by the nonreal complex dimensions $\omega$ and $\overline{\omega}$ of $\mcl$ remain (as is obvious from \eqref{6.3}), but due to the fact that by construction, $\zeta (\omega) = \zeta (\overline{\omega}) = 0$, the spectral oscillations (in the asymptotic second term of $N_\nu (x)$) disappear. Therefore, we see the subtle interplay between complex dimensions, geometric and spectral oscillations, as well as the critical zeros of 
$\zeta$. (Clearly, both $\omega$ and $\overline{\omega}$ belong to the open critical strip $0 < \Res < 1$ since we have that $\text{Re}(\omega) = \text{Re}(\overline{\omega}) =D$ and $D \in (0,1)$.) \\

\tab Finally, we note that by using the theory of complex dimensions developed in [Lap-vFr1--3], it can be shown that the above fractal $\mcl$ has exactly three complex dimensions (each of which has a real part equal to $D$ and multiplicity one). Namely, the set $\mathcal{D}_\mcl$ of complex dimensions of $\mcl$ is given by $\mathcal{D}_\mcl = \{D, \omega, \overline{\omega} \}$. Moreover, observe that $x^\omega = x^D \ x^{i \tau}$ and $x^\omega = x^D \ x^{-i \tau}$; so that the real part (resp., the imaginary part) of the complex dimension $\omega$ (or $\overline{\omega}$) determines the {\em amplitude} (resp., the {\em frequency}) of the corresponding geometric or spectral oscillations (viewed multiplicatively). This statement is now fully corroborated (for any fractal string and its complex dimensions) by the rigorous theory of complex dimensions of fractal strings and the associated (generalized) explicit formulas developed in [Lap-vFr1--3].

\section{Epilogue: Later Developments and Research Directions}
In this epilogue, by necessity of concision, we very briefly discuss further developments closely connected to (or partly motivated by) the results discussed in the main body of this paper as well as by related results and conjectures in [Lap1--3]. These topics include a geometric interpretation of the critical strip for the Riemann zeta function $\zeta = \zeta(s)$ (see \S 7.1), the theory of complex dimensions of fractal strings (see \S 7.2), fractal zeta functions and a higher-dimensional theory of complex dimensions (valid for arbitrary bounded subsets of Euclidean spaces and relative fractal drums, see \S 7.3), quantized number theory and the spectral operator, along with a functional analytic reformulation of the results of \cite{LapMai2} discussed in \S6, as well as a different framework (developed in [HerLap1--5]) and a new asymmetric reformulation of the Riemann hypothesis recently obtained by the author in \cite{Lap7} (see \S 7.4).

\subsection{Fractal strings, $\zeta = \zeta(s)$, and a geometric interpretation of the critical strip}
The one-dimensional situation (i.e., the case of fractal strings) is ideally suited to the Riemann zeta function $\zeta = \zeta(s)$ in the (closed) critical strip $0 \leq \Res \leq 1$, as we have seen in \S5 and, especially, \S6 above. This is due in part to the product formula \eqref{1.1}, $\zeta_\nu (s) = \zeta (s) \cdot \zeta_\mcl (s)$, connecting the geometric zeta function $\zeta_\mcl$ of a fractal string $\mcl$ to its spectral zeta function 
$\zeta_\nu$,\footnote{We note that the product formula \eqref{1.1} for the spectral zeta functions of fractal strings has since been extended in various ways in the setting of Laplacians on certain self-similar fractals; see [Tep1--2, DerGrVo, LalLap1--2].}  
combined with the fact that fractal strings always have a (Minkowski) dimension between $0$ and $1: 0 \leq D \leq 1$.\\


\tab These facts, along with the results of [LapPom1,2] and some of the results and conjectures of [Lap1--3], have led to the following geometric interpretation of the (closed) critical strip $0 \leq \Res \leq 1$, using the terminology introduced in \cite{Lap1} (in any dimension): The least (resp., most) fractal case when $D=0$ (resp., $D=1$), corresponds to the left- (resp., right-) hand side of the critical strip, that is, to the vertical line $\{\Res =0 \}$ (resp., $\{\Res =1 \}$. Furthermore, the midfractal case when $D=1/2$ corresponds to the critical line, namely, the vertical line $\{\Res = 1/2 \}$ where (according to the Riemann hypothesis) all of the nontrivial (or critical) zeros of $\zeta = \zeta(s)$ are supposed to be located. This geometric picture of the critical strip has later been corroborated by the work in [LapMai1,2] described in \S6 above. Its complete and rigorous justification has then been provided by the theory of complex dimensions for fractal strings (that is, in the one-dimensional situation) developed in [Lap-vFr1--3]. (See, in particular, [Lap-vFr3, Chapters 9 and 11].) We will next briefly describe (in \S 7.2) a few aspects of the latter theory.

\subsection{Complex dimensions of fractal strings and oscillatory phenomena}
The theory of complex dimensions of fractal strings, developed by the author and M. van Frankenhuijsen in the research monographs [Lap-vFr1--3] (and several corresponding articles), aimed originally at obtaining a much more accurate understanding of the oscillatory phenomena which are intrinsic to fractals (in their geometries and their spectra as well as in the underlying dynamics). This is accomplished via explicit formulas (generalizing Riemann's original 1858 explicit formula and its extensions to various aspects of number theory and arithmetic geometry, see [Edw, Ing, ParSh1--2, Pat, Sarn, Den1, Ti] along with [Lap-vFr3, \S 5.1.2 and \S 5.6]) expressed in terms of the underlying complex dimensions; see [Lap-vFr3, Chapter 5]. The latter complex dimensions are defined as the poles of a suitable zeta function, typically a geometric, spectral, dynamical or arithmetic zeta function. (See \cite{Lap-vFr3}.)\\

\tab In the case of fractal strings, which is the main focus of the theory developed in [Lap-vFr1--3], these explicit formulas can be applied, for example, to the geometric counting function $N_\mcl (x)$, the spectral counting function $N_\nu (x)$, the volume of tubular neighborhoods $V(\varepsilon)$ (giving rise to a ``fractal tube formula''), or to the counting function of the number of primitive geodesics of an underlying dynamical system. (See [Lap-vFr3, Chapters 6--11].) In the special case of self-similar strings, the resulting fractal tube formulas give very precise information, in part due to the knowledge of the periodic (or, in general, the quasiperiodic) structure of the complex dimensions (see [Lap-vFr3], Chapters 2--3 and \S 8.4). \\

\tab For a fractal string $\mcl$, under mild growth assumptions on the associated geometric zeta function $\zeta_\mcl$ (see [Lap-vFr3, \S 5.1]) and assuming that all the complex dimensions (i.e., the poles of $\zeta_\mcl$) are 
simple, the {\em fractal tube formula} for $V(\varepsilon) = V_\mcl (\varepsilon)$ takes the following form (see [Lap-vFr3, 
Chapter 8]):

\begin{equation}\label{7.1}
V(\varepsilon) = \sum_{\omega \in \mathcal{D}} \alpha_\omega \frac{ (2 \varepsilon)^{1-\omega}}{\omega (1-\omega)} + R(\varepsilon),
\end{equation}
where $\mcd = \mcd_\mcl$ denotes the set of (visible) complex dimensions of $\mcl$ and $\alpha_\omega := res (\zeta_\mcl, \omega)$ for each $\omega \in 
\mcd$. Furthermore, the error term $R (\varepsilon)$ is precisely estimated in \cite{Lap-vFr3}. In the important special case of self-similar strings (which includes the Cantor string discussed just below), one can take $R(\varepsilon) \equiv 0$ and therefore obtain an {\em exact} fractal tube formula (which holds pointwise); see [Lap-vFr3, \S 8.4].\\

\begin{rem}\label{R6.45}
$($a$)$ The explicit formula is valid for multiple poles as well but must then take a different form, in general. Namely, we then have $($see $[$Lap-vFr3, Theorems 8.1 and 8.7$])$
\begin{equation}\label{29.5}
V(\varepsilon) = \sum_{\omega \in \mcd} res \left(  \frac{\zeta_\mcl (s) \ (2\varepsilon)^{1-s}}{s (1-s)}, \omega \right) + R(\varepsilon).
\end{equation}\\

$($b$)$ The fractal tube formulas \eqref{7.1} and \eqref{29.5} hold pointwise or distributionally, depending on the growth assumptions made about $\zeta_\mcl$; see $[$Lap-vFr3$]$, \S 8.1, especially, Theorems 8.1 and 8.7 and their corollaries. Also, in the so-called ``strongly languid'' case $($in the sense of $[$Lap-vFr3, Definition 5.3$])$, we can let $R (\varepsilon) \equiv 0$ and therefore obtain an {\em exact} formula. This is the case, for example, for all self-similar strings$;$ see $[$Lap-vFr3, \S 8.4$]$.\\

$($c$)$ For the first higher-dimensional analog of \eqref{7.1} and \eqref{29.5}, see $[$Lap-vFr1--3$]$ for specific examples of fractal sprays $($in the sense of $[$LapPom3$])$, and for a fairly general class of fractal sprays and self-similar tilings $($or, less generally, sets$)$, see $[$LapPe2--3, LapPeWi1--2$];$ see also $[$Pe, PeWi$]$ and, for a direct approach in the case of the Koch snowflake curve, $[$LapPe1$]$. Within the general higher-dimensional theory of complex dimensions developed in $[$LapRa\u Zu1--8$]$, the precise counterpart of \eqref{7.1} and \eqref{29.5} is provided in $[$LapRa\u Zu5--6$]$ and $[$LapRa\u Zu1, Chapter 5$];$ see \S 7.2 for a brief discussion.
\end{rem}

\tab Let us now specialize the above discussion to the Cantor string $\mcl = CS$, viewed geometrically as the bounded open set $\Omega \subset \mbr$, defined as the complement of the classic ternary Cantor set in $[0,1]$; hence, $\partial \Omega$ is the Cantor set. Then $\mcl = CS = (\ell_j)_{j=1}^\infty$, with $\ell_1 = 1/3, \ell_2 = \ell_3 = 1/9, \ell_4 = \ell_5 = \ell_6 = \ell_7 = 1/27, \cdots .$ Alternatively, the lengths of the Cantor string (or harp) are the numbers $3^{-n-1}$ repeated with multiplicity $2^n$, for $n = 0,1,2, \cdots$. Then

\begin{equation}\label{7.2}
\zeta_{CS} (s) = \frac{3^{-s}}{1-2 \cdot 3^{-s}}, \text{ for all } s \in \mathbb{C};
\end{equation}
so that the set $\mcd_{CS}$ of complex dimensions (here, the complex solutions of the equation $1-2 \cdot 3^{-s} = 0$) is given by

\begin{equation}\label{7.3}
\mcd_{CS} = \{D + in{\bf p} : n \in \mathbb{Z} \},
\end{equation}
where $D := \log_3 2 = \log 2/ \log 3$ (the Minkowski dimension of $CS$ or, equivalently, of the ternary Cantor set) and ${\bf p} := 2 \pi/\log 3$ (the {\em oscillatory period} of $CS$). Then \eqref{7.1} (with $R(x) \equiv 0$) becomes

\begin{align}\label{7.4}
V_{CS} (\varepsilon) &= \frac{1}{2 \log 3} \sum_{n \in \mathbb{Z}} \frac{(2 \varepsilon)^{1-D-in{\bf p}}}{(D + in{\bf p})(1 - D - in{\bf p})} - 2 \varepsilon \\
\notag &= \varepsilon^{1-D} \ G(\log \varepsilon^{-1}) - 2 \varepsilon,
\end{align}
where $G$ is a periodic function which is bounded away from zero and infinity. One then recovers the result from [LapPom1,2] according to which $CS$ (and hence also, the Cantor set) is Minkowski nondegenerate  but is not Minkowski measurable. (Actually, the values of $\mcm_*$ and $\mcm^*$ are computed explicitly in \cite{LapPom2} as well as, in a more general context, in [Lap-vFr3, Chapter 10].) \\

\tab Moreover, the geometric counting function of $CS$ is given by

\begin{equation}\label{7.5}
N_{CS} (x) = \frac{1}{2 \log 3} \sum_{n \in \mathbb{Z}} \frac{x^{D + in{\bf p}}}{D + in{\bf p}} - 1
\end{equation}
while the corresponding frequency counting function is given by

\begin{equation}\label{7.6}
N_{\nu, CS} (x) = x + \frac{1}{2 \log 3} \sum_{n \in \mathbb{Z}} \zeta (D + in{\bf p}) \frac{x^{D + in{\bf p}}}{D + in{\bf p}} + O(1) \text{ as } x \rightarrow + \infty.
\end{equation}
(See \cite{Lap-vFr3}, \S 1.1.1, \S 1.2.2 and Chapter 6.)\\

\tab In \eqref{7.1} and \eqref{7.4}--\eqref{7.6}, we see the intuitive (and actual) meaning of the complex dimensions. Namely, the real (resp., imaginary) parts correspond to the {\em amplitudes} (resp., {\em frequencies}) of the oscillations (in the spaces of scales, for \eqref{7.1} and \eqref{7.4}--\eqref{7.5}, and in frequency space, for \eqref{7.6}).

\begin{rem}[Fractality and complex dimensions]\label{R7.1}
The notion of fractality is notoriously difficult to define. Mandelbrot $\cite{Man}$ has proposed to define a fractal as a geometric object whose Hausdorff dimension is strictly greater than its topological dimension. There is an obvious problem with this definition $($Mandelbrot was aware of it, as is stated in his book, $[$Man, p. 82$])$. Namely, the Cantor curve $($or ``devil's staircase''$)$ is not fractal in this sense $($since its Hausdorff, Minkowski and topological dimensions are all equal to 1$);$ however, as Mandelbrot states in $\cite{Man}$, everyone would agree that the Cantor curve should be called ``fractal''. This issue has long preoccupied the present author.\\
\indent \indent There is, however, a satisfactory way to resolve this apparent paradox as well as many other related and unrelated issues. In the theory of complex dimensions developed in $[$Lap-vFr1--3$],$ an object is called ``fractal'' if it has at least one {\em nonreal} complex dimension $($with positive real part$)$. $($See $[$Lap-vFr3, \S 12.1$].)$ Accordingly, the Cantor curve $($CC$)$ is, indeed, fractal because its set of complex dimensions is given by $\mcd_{CC} = \mcd_{CS} \cup \{ 1\}$, where $\mcd_{CS}$ is the set of complex dimensions of the Cantor set $($or, equivalently, the Cantor string$)$ given by \eqref{7.3}. $($Hence, $\mcd_{CC}$ has infinitely many nonreal complex dimensions with positive real part.$)$ Furthermore, $($nontrivial$)$ self-similar strings $($and, more generally, self-similar geometries) are fractal.\footnote{An example of a trivial self-similar set is an interval of $\mbr$ or a cube in $\mbr^N (N \geq 2).$ In such cases, all of the complex dimensions are easily seen to be real; see \cite{Lap-vFr3} and \cite{LapRaZu1}.}
 In order to accommodate random $($and, in particular, stochastically self-similar fractals$)$, as in $\cite{HamLap}$ and the relevant references therein, the above definition of fractality has been extended to allow for a ``fractal'' $A$ to be such that its associated zeta function $\zeta_A$ has a natural boundary along a suitable curve of the complex plane, called a ``screen'' in $[$Lap-vFr3, \S 5.1$]$ $($and hence, such that $\zeta_A$ cannot be meromorphically extended beyond that curve$)$. In $[$LapRa\u Zu1--8$]$, such a compact subset of $\mathbb{R}^N$ is called a ``hyperfractal''. In $[$LapRa\u Zu1,2,4$]$ are even constructed ``maximal hyperfractals''$;$ that is, compact subsets $A$ of $\mathbb{R}^N$ $($where $N \geq 1$ is arbitrary$)$ such that $\zeta_A$ has a singularity at every point of the ``critical line'' $\{\Res =D \}$, where $D$ is the Minkowski dimension of $A$.
\end{rem}

\begin{rem}[Noncommutative Riemann flow of zeros and the Riemann hypothesis, \cite{Lap6}]\label{R7.2}
In the long term, the theory of complex dimensions aims at unifying many aspects of fractal and arithmetic geometries. Many concrete and more abstract examples are provided in $\cite{Lap-vFr3}.$ This direction is pursued in many different directions in the author's book, {\em In Search of the Riemann Zeros} $\cite{Lap6}$, where is introduced the notion of fractal membrane $($i.e., quantized fractal string$)$ and the associated moduli spaces of fractal membranes $($as well as of fractal strings$)$. In $\cite{Lap6}$, a still conjectural $($noncommutative$)$ flow on the moduli space of fractal membranes and correspondingly, a flow of zeta functions $($or partition functions$)$ and a flow of zeros are used in an essential manner in order to provide a possible new interpretation of $($and approach to$)$ the Riemann hypothesis. \\
\indent Accordingly, the flow of fractal membranes can be viewed as some kind of noncommutative Ricci flow which is transforming $($as ``time'' tends to infinity or physically, as the absolute ``temperature'' tends to zero$)$ generalized quasicrystals into $($self-dual$)$ pure crystals. Correspondingly, the zeta functions are becoming increasingly symmetric $($that is, satisfy a true functional equation, in the limit$)$, while the zeros $($of these zeta functions$)$ are pushed under this flow $($viewed as a flow on the Riemann sphere$)$ onto the Equator, which naturally represents the critical line in this context. Conjecturally, this would also provide a proof of the generalized Riemann hypothesis $($GRH$)$, valid for all of the number theoretic zeta functions for which the analog of the Riemann hypothesis is expected to be true. $($See, e.g., $[$ParSh1--2$]$, $[$Sarn$]$, $[$Lap-vFr3, Appendix A$]$ and $[$Lap6, Appendices B, C and E$].)$
Very concisely, this is the interpretation of $($and approach to$)$ the Riemann hypothesis proposed in $\cite{Lap6}$. Needless to say, it poses formidable mathematical and physical challenges but may nevertheless stimulate further investigations of the aforementioned $($noncommutative$)$ Riemann $($or ``modular''$)$ flow, whether it be viewed as a $($noncommutative$)$ Ricci-type flow or, physically, as a renormalization flow $($not unlike the one which presumably describes the time evolution of our universe$)$.
\end{rem}

\subsection{Fractal zeta functions of arbitrary compact sets and higher-dimensional theory of complex dimensions}
In a forthcoming book by the author, Goran Radunovi\' c and Darko \u Zubrini\' c, entitled {\em Fractal Zeta Functions and Fractal Drums}, \cite{LapRaZu1} (see also the series of research and survey papers [LapRa\u Zu2--8]), is developed a higher-dimensional theory of complex dimensions, valid for any bounded subset of $N$-dimensional Euclidean space $\mathbb{R}^N$, with $N \geq 1$ arbitrary. Distance and tube zeta functions are defined in that general setting. Let $D$ be the (upper) Minkowski dimension of a given (nonempty) bounded subset $A$ of $\mathbb{R}^N$. Then it is shown in [LapRa\u Zu1--2] that the {\em distance zeta function} of $A$, denoted by $\zeta_A = \zeta_A (s)$ and defined (for all $s \in \mathbb{C}$ with $\Res$ sufficiently large and for a given $\delta > 0$) by 

\begin{equation}\label{7.7}
\zeta_A (s) := \int_{A_\delta} d (x, A)^{s-N} dx
\end{equation} 
is holomorphic for 
$\Res > D$.\\

\begin{rem}\label{R17.55}
$($a$)$ This new type of fractal zeta function, $\zeta_A$, was introduced by the author in 2009 in order to be able to extend to any higher dimension $N$ the theory of complex dimensions of fractal strings developed in $[$Lap-vFr1--2$]$ $($and now also in $[$Lap-vFr3$])$. The case of fractal strings corresponds to $N=1;$ in that case, the precise relationship between $\zeta_\mcl$ and $\zeta_A$, with $A:= \partial \mcl$, is explained in $[$LapRa\u Zu1$]$ and $[$LapRa\u Zu2$]$. \\

$($b$)$ For the simplicity of the discussion, we assume here that $|A|=0$ $($i.e., $A$ is a Lebesgue null set$)$, which is the case of most fractals of interest. We refer to $[$LapRaZu1$]$ for a discussion of the general case.

\end{rem}

\tab A first key result of the theory is that the abscissa of convergence of this Dirichlet-type integral (viewed as a Lebesgue integral) or, equivalently, the {\em abscissa of} ({\em absolute}) {\em convergence} of $\zeta_A$, coincides with $D$: 

\begin{equation}\label{7.8}
\sigma = D,
\end{equation}
the (upper) Minkowski (or box) dimension of 
$A$. Therefore, $\{\Res > D \}$ is the largest open right half-plane (of the form $\{ \Res > \alpha \}$, for some $\alpha \in \mbr \cup \{\pm \infty \}$), on which the Lebesgue integral appearing on the right-hand side of \eqref{7.7} is convergent (i.e., absolutely convergent). This result is the higher-dimensional counterpart of Theorem \ref{T2.1} above (first noted in [Lap2--3] in the case of fractal strings, i.e., when $N=1$). The proof of Equation \eqref{7.8} makes use of an interesting integral estimate obtained in \cite{HarvPol} in order to study the singularities of solutions of certain linear partial differential equations. (See also \cite{Zu1}.)\\

\tab Moreover, under mild assumptions, namely, if 
$D < N$ (recall that we always have $0 \leq D \leq N$), $\mcm_*^D > 0$ and the Minkowski dimension of $A$ exists (i.e.,  the lower and upper Minkowski dimensions of $A$ coincide), then $\{\Res > D \}$ is also the maximal open right-half plane to which $\zeta_A = \zeta_A(s)$ can be holomorphically continued; i.e., $D$ coincides with $D_{hol} (\zeta_A),$ the {\em abscissa of holomorphic continuation} of $\zeta_A$:

\begin{equation}\label{7.9}
D = \sigma = D_{hol} (\zeta_A),
\end{equation}
where, as above, $D=D(A)$ and $\sigma = D_{abs} (\zeta_A)$.\\

\tab An entirely analogous theorem holds for the {\em tube zeta function} of $A$, denoted by $\zeta_A = \widetilde{\zeta}_A (s)$ and defined by

\begin{equation}\label{7.10}
\widetilde{\zeta}_A (s) = \int_0^\delta \ t^{s-N} |A_t|_N \frac{dt}{t},
\end{equation}
for all $s \in \mbc$ such that $\Res > D$.\\

\tab In fact, it can be shown that the two functions, $\zeta_A$ and $(N-s) \ \widetilde{\zeta}_A$, differ by an entire function, from which it follows that (for $D <N$), the two fractal zeta functions $\zeta_A$ and $\widetilde{\zeta}_A$ have exactly the same qualitative properties. In particular, given a domain $U \subseteq \mathbb{C}$ containing $\{\Res > D \}$, the common half-plane of (absolute) convergence of $\zeta_A$ and $\widetilde{\zeta}_A$, $\zeta_A$ can be meromorphically extended to $U$ if and only if it is the case for $\widetilde{\zeta}_A$. Furthermore, $\zeta_A$ and $\widetilde{\zeta}_A$ have exactly the same poles, in $U$ (with the same multiplicities). These poles are called the (visible) {\em complex dimensions} of $A$. Moreover, the residues of $\zeta_A$ and $\widetilde{\zeta}_A$ at a simple pole (or, more generally, the principal parts at a multiple pole) are related in a very simple way. For example, if $D$ is simple, then we have \[res(\zeta_A, D) = (N-D) \ res(\widetilde{\zeta}_A,D).\]

\tab Finally, if $\zeta_A$ (or, equivalently, $\widetilde{\zeta}_A$) can be meromorphically continued to an open and connected neighborhood of $D$, then $D$ is a simple pole of $\zeta_A$ and $\widetilde{\zeta}_A$ (i.e., it is a complex dimension of 
$A$)\footnote{Since $\zeta_A$ and $\widetilde{\zeta}_A$ are holomorphic in the open half-plane $\{\Res > D \}$, it then follows that

\begin{equation}\label{7.11}
D = \max \{\Res: s \in \mathcal{D} \},
\end{equation}
where $\mcd$ is the set of (visible) complex dimensions of $A$ in any domain $U$ containing this neighborhood of $\{D \}$.}
and the residue of $\widetilde{\zeta}_A$ at $s =D$ is squeezed between the lower and upper Minkowski contents of $A$:

\begin{equation}\label{7.12}
\mcm_* \leq res (\widetilde{\zeta}_A; D) \leq \mcm^*.
\end{equation}
If, in addition, $A$ is Minkowski measurable, (with Minkowski content denoted by $\mcm$, as before), then

\begin{equation}\label{7.13}
res (\widetilde{\zeta}_A; D) = \mcm.
\end{equation}

\tab Both formulas \eqref{7.12} and \eqref{7.13} are valid even if $D=N$, although the justification of this statement requires a specific argument; see \cite{LapRaZu1}.\\

\tab We should note that, in light of the aforementioned functional equation connecting $\zeta_A$ and $\widetilde{\zeta}_A$, the choice of $\delta$ is unimportant in the definition of $\zeta_A$ and $\widetilde{\zeta}_A$ (in \eqref{7.7} and \eqref{7.10}, respectively). Furthermore, the residues of $\zeta_A$ and $\widetilde{\zeta}_A$ at (simple) complex dimensions are independent of $\delta$.\\

\tab In the special case when $N=1$ (that is, for fractal strings), it is shown (still in [LapRa\u Zu1,2,8]) that the distance zeta function $\zeta_A$ and the geometric zeta function $\zeta_\mcl$ are closely related (here, $A := \partial \Omega$ and $\mcl$ is the fractal string associated with the bounded open set $\Omega \subset \mathbb{R}$). In fact, given any subdomain $U$ of $\{\Res > 0 \}$ (or, more generally, of $\mathbb{C} \backslash \{0 \}), \zeta_\mcl$ has a meromorphic continuation to $U$ if and only if $\zeta_A = \zeta_{\partial \Omega}$ does. In fact, their difference can be shown to be holomorphic in $U$. Consequently, $\zeta_\mcl$ and $\zeta_A$ have the same poles (with the same multiplicities) in $U$; i.e., they have the same visible complex dimensions. This is true independently of the geometric realization $\Omega$ of the fractal $\mcl = (\ell_j)_{j=1}^\infty$ as a bounded open subset of $\mathbb{R}$. In particular, we recover Theorem \ref{T2.1}.\\

\tab A variety of results are provided in [LapRa\u Zu1--8], guaranteeing the existence of a suitable meromorphic extension of $\zeta_A$ (and of $\widetilde{\zeta}_A$) beyond the half-plane of convergence $\{\Res > D \}$. Moreover, many examples of computation of the complex dimensions of a variety of fractals (when $N=1, 2$ or $N \geq 3$) are provided throughout [LapRa\u Zu1--8]. These fractals include (one- or two-parameter) families of generalized Cantor sets, the Sierpinski gasket and carpet, the Menger sponge (a 3-dimensional analog of the Sierpinski carpet, see, e.g., \cite{Man}), families of spirals, fractal curves such as the Cantor curve (or devil's staircase), which is examined from several points of view, as well as geometric chirps and curves with cusps. The theory is extended to a new class of objects, called relative fractal drums 
(RFDs), which allow a much greater flexibility in many situations and generalize both the usual fractal drums (i.e., drums with fractal boundary, in the sense, for example, of [Lap1--3], as discussed in \S4 above) and the class of bounded subsets of Euclidean space $\mathbb{R}^N$. It is noteworthy that in the case of RFDs, the relative Minkowski dimension can be negative; it can even take the value $- \infty$, as is shown in [LapRa\u Zu1,4,7], where geometric explanations are provided for this ``dimension drop'' phenomenon.\\

\tab In short, a {\em relative fractal drum} is a pair $(A, \Omega),$ with $A$ an arbitrary (possibly unbounded) subset of $\mathbb{R}^N$ and $\Omega$ an open subset of $\mathbb{R}^N$ with finite volume (i.e., $|\Omega|_N < \infty$) and such that $\Omega \subseteq A_\delta$, for some $\delta > 0.$ The special case of an ordinary fractal drum (or ``drum with fractal boundary'', in the sense of [Lap1--3]) corresponds to the case where $A = \partial \Omega$ and $\Omega$ is as above.\\

\tab Furthermore, the main results of \cite{Lap1} (see Section 4.1 above, especially, Theorem \ref{T4.1}) are used in an essential manner in order to show that the spectral zeta function of a fractal drum (that is, of the Dirichlet Laplacian on a bounded open subset of $\mbr^N,$ with $N \geq 1$) can be meromorphically extended to (at least) the open half-plane $\{\Res > D \}$, where (as before) $D$ is the (upper) Minkowski dimension of the boundary $\partial \Omega$ of the drum. [This was already observed in \cite{Lap3}, using \cite{Lap1} and a well-known Abelian-type theorem (in the sense of \cite{Sim}), the converse of a Tauberian theorem. It can also be deduced from the main result of \cite{Lap1} by a direct argument, based on the holomorphicity of an integral depending analytically on a parameter.] Moreover, using the construction of a maximal 
hyperfractal\footnote{Recall that in the terminology of [LapRa\u Zu1--8], a ``maximal hyperfractal'' is a compact subset $A$ of $\mathbb{R}^N$ such that $\zeta_A$ has singularities at every point of the ``critical line'' $\{\Res =D \}$; in addition to Remark \ref{R7.1}, see Remark \ref{R13.5} below.}
carried out in [LapRa\u Zu1,2,7] (and alluded to towards the end of Remark \ref{R7.1} above), it is shown that the half-plane $\{\Res > D \}$ is optimal, in general; i.e., it cannot usually be larger, from the point of view of the meromorphic continuation of the spectral zeta function. We note that under suitable assumptions, the Dirichlet Laplacian can be replaced by a higher order positive, self-adjoint elliptic operator, with possibly variable coefficients and with Dirichlet, Neumann or mixed boundary conditions; see \cite{Lap1} and [LapRa\u Zu1,7].\\

\tab Extensions of the results of [LapRa\u Zu1--8] to unbounded sets, in particular, are provided in \cite{Ra}.\\

\tab Finally, we note that $\zeta_A$ and $\widetilde{\zeta}_A$ remain unchanged if we replace the bounded set $A \subset \mathbb{R}^N$ by its closure $\overline{A}$. As a result, we could have assumed throughout that $A$ was a compact subset of $\mathbb{R}^N$. Moreover, we point out that fractal tube formulas, significantly extending to compact subsets of $\mathbb{R}^N$ the corresponding (pointwise and distributional) tube formulas obtained in [Lap-vFr3, \S 8.1--8.3] for fractal strings (as well as the later tube formulas obtained for fractal sprays in [LapPe2--3] and [LapPeWi1--2]), are established in [LapRa\u Zu5--6] and [LapRa\u Zu1, Chapter 5], without any assumptions of self-similarity and in every dimension $N \geq 1$.

\begin{rem}\label{R13.5}
We close this discussion by mentioning the fact that the higher-dimensional theory of fractal zeta functions and the associated complex dimensions now enables us to extend to the general $($higher-dimensional$)$ setting the definition of fractality introduced in $[$Lap-vFr1--3$]$ $($see, especially, $[$Lap-vFr3, \S 13.1 and \S 13.2$])$. Accordingly, a geometric object is said to be ``fractal'' if it has at least one {\em nonreal} complex dimension $($with a positive real part$)$. In particular, one can now precisely talk about the fractality of any bounded subset $A$ $($or, more generally, relative fractal drum$)$ in $\mbr^N$, with $N \geq 1$ arbitrary. $($Compare with Remark \ref{R7.1} above.$)$\\
\indent \indent As an example, the Cantor curve $($i.e., the ``devil's staircase'' in the terminology of $\cite{Man})$ is fractal according to this definition, whereas it is not fractal according to Mandelbrot's original definition in $\cite{Man}$ $($because its Hausdorff, box and topological dimensions coincide and are equal to one$)$. $($Recall from Remark \ref{R7.1} that in $[$Man$]$, a geometric object is said to be ``fractal'' if its Hausdorff dimension is strictly greater than its topological dimension.$)$ One can also check that most of the classic ``fractals'' $($for example, the ternary Cantor set and its generalizations, as well as the Sierpinski gasket and carpet and their higher-dimensional counterparts$)$ are indeed fractal in this new sense. $($See $[$LapRa\u Zu1--4,7$]$ for these and many other examples.$)$\\
\indent \indent A technical challenge remains to prove this same result for all $($suitable$)$ self-similar sets $($as was done when $N=1$ for self-similar strings in $[$Lap-vFr1--3$]$ and when $N \geq2$ for a large class of self-similar sprays or tilings in $[$LapPe2--3$]$ and 
$[$LapPeWi1--2$])$,\footnote{{\em Added note}: In the case of self-similar sprays, the results of [LapRa\u Zu1,5,6] now enable one to recover and significantly extend the results of [LapPe2--3] and [LapPeWi1,2].} 
as well as for the classic types of fractals occurring in complex dynamics $($Julia sets, the Mandelbrot set and their generalizations$;$ see, e.g., $\cite{Man, TanL, Shi})$ and in conformal dynamics and/or hyperbolic geometry $($for instance, limit sets of Fuchsian groups and Kleinian groups; see, e.g., $\cite{BedKS})$, possibly by using other gauge functions than the usual ones based on the standard power laws, as was originally done in $\cite{HeLap}$ and further studied in $[$LapRa\u Zu1--8$]$ $($from the point of view of the new higher-dimensional theory of complex dimensions$)$. It may be difficult to do so, but there is no doubt in the author's mind that all of the classic $($deterministic$)$ fractals will eventually be found to be ``fractal'' in the above sense or ``hyperfractal'', in a sense to be explained next.\\
\indent \indent Due in part to the work in $\cite{HamLap}$ on random fractal strings and their complex dimensions, the notion of fractality was extended as follows $($first in $\cite{Lap-vFr2}$ and $[$Lap-vFr3, \S 13.4.3$]$ and then, in any dimension, in $[$LapRa\u Zu1--8$])$. A geometric object is said to be ``fractal'' if it has at least one nonreal complex dimension with a positive real part $($as above$)$ or else if it is ``hyperfractal''$;$ i.e., if the associated fractal zeta function has a natural boundary along some suitable contour in $\mbc$ $($a ``screen'', in the sense of $[$Lap-vFr3, \S 5.3$])$. We note that the term ``hyperfractal'' was introduced in $[$LapRa\u Zu1--8$]$, in this context.\\
\indent \indent Therefore, a ``hyperfractal'' is such that the associated fractal zeta function cannot be meromorphically extended beyond a certain ``screen''. Furthermore, in $[$LapRa\u Zu1--4,7--8$]$ is also introduced the notion of ``maximal hyperfractal'', according to which the corresponding fractal zeta function has a singularity at every point of the critical line of convergence $\{\Res =D \}$, where $D$ is the Minkowski dimension of $A \subset \mbr^N$ $($or its relative counterpart, in the case of an RFD$)$. It is then shown in $([$LapRa\u Zu2--4,7$]$ or $[$LapRa\u Zu1, Chapter 4$])$ using, in particular, countably many suitable fractal strings assembled in an appropriate way, along with Baker's theorem from transcendental number theory $\cite{Bak}$, that maximally hyperfractal strings $(N=1)$, as well as maximally hyperfractal compact subsets and RFDs of $\mbr^N$ $($for every $N \geq 1)$ can be explicitly constructed. This construction is completely deterministic. The author conjectures $($building on $\cite{HamLap}$ and $[$Lap-vFr3, \S13.4.3$])$ that for large classes of random fractals, maximal hyperfractality is an almost sure property.\\
\indent \indent In closing this remark, we mention that the above definition of fractality cannot just be applied to standard geometric objects embedded in Euclidean spaces $($or, more generally, in appropriate metric measure spaces$)$ but is also applicable, in principle, to `virtual' geometries, spectral geometries, dynamical systems, algebraic and noncommutative geometries $($not necessarily consisting of ordinary points$;$ see, e.g., $\cite{Cart}$ and $\cite{Con})$, as well as arithmetic geometries. In fact $($along with the proper notion of zeta function and the associated complex dimensions$)$, it should be used as a unifying tool between these apparently vastly different domains of mathematics. This long-term goal has been one of the central motivations of the author $($and his collaborators$)$ in $[$Lap1--9, Lap-vFr1--3, LapRa\u Zu1, LapPom2, LapMai2, HeLap, LapLu3, LapNe, LapPeWi1, EllLapMaRo, ChrIvLap, LapSar$]$ $($particularly, in $[$Lap3--5$]$ and $[$Lap8--9$]$, as well as in the books $\cite{Lap6}$ and $[$Lap-vFr1--3$])$. \\
\indent \indent For instance, the author conjectures that there exists a natural fractal-like geometric object whose complex dimensions are precisely the critical $($i.e., nontrivial$)$ zeros of the Riemann zeta function. Furthermore, the essential ``shape'' of this object should be understood in terms of a yet to be constructed cohomology theory associated with the underlying complex dimensions $($and the pole of $\zeta = \zeta(s)$ at $s=1)$. In particular, connections with Deninger's work $($and conjectures$)$ in $[$Den1,2$]$ arise naturally in this context. $($See, especially, $[$Lap6$]$ and $[$Lap8--9$]$, along with $[$Lap-vFr3, \S 12.3 and \S 12.4$].)$ 
\end{rem}

\subsection{Quantized number theory, spectral operators and the Riemann hypothesis.}
Formula \eqref{1.1}, $\zeta_\nu (s) = \zeta (s) \cdot \zeta_\mcl (s)$, which connects the spectral zeta function $\zeta_\nu = \zeta_{\nu, \mcl}$ and the geometric zeta function $\zeta_\mcl$ of a fractal string $\mcl$ via the Riemann zeta function $\zeta$, has the following counterpart, in terms of the spectral and geometric counting functions of $\mcl$ (see [Lap-vFr3, Theorem 1.2]):

\begin{equation}\label{7.14}
N_\nu (x) = \sum_{n=1}^\infty N_\mcl \left( \frac{x}{n} \right),
\end{equation}
for any $x > 0$. Note that for a fixed $x > 0$, only finitely many terms are nonzero on the right-hand side of \eqref{7.14}. However, the number of these terms tends to $+ \infty$ as $x \rightarrow + \infty$.\\

\tab The ({\em heuristic}) {\em spectral operator}, at the level of the counting functions, is then given by the map

\begin{equation}\label{7.15}
g := N_\mcl \mapsto N_\nu (g) := \sum_{n=1}^\infty g \left( \frac{\cdot}{n} \right);
\end{equation}
that is, $N_\nu (g)(x) := \sum_{n=1}^\infty g(x/n)$, for all $x>0$. It can therefore be thought of as the map sending the geometry (represented by $N_\mcl$, given by Equation \eqref{3.4}) of a fractal string $\mcl$ onto the spectrum (represented by $N_\nu = N_{\nu, \mcl}$) of a fractal string $\mcl$. Here, as in the discussion preceding Equation \eqref{3.6},

\begin{equation}\label{7.16}
N_\nu (x) = \# (\{f \in \sigma (\mcl): f \leq x \}),
\end{equation}
for any $x > 0$, where $\sigma (\mcl) = \{n \cdot \ell_j^{-1}: n \geq 1, j \geq 1 \}$ is the (frequency) spectrum of $\mcl = 
(\ell_j)_{j=1}^\infty$.

\begin{rem}\label{R18.5}
At an even more fundamental level $($that of the ``density of geometric states'' $\eta := \sum_{j=1}^\infty \delta_{\{\ell_j^{-1} \}}$ and the ``density of spectral states'' $\nu := \sum_{f \in \sigma (\mcl)} \delta_{\{ f\}}$, see $[$Lap-vFr3, \S 6.3.1$])$, the spectral operator can be viewed as the map sending the geometry $($represented by $\eta)$ of a fractal string $\mcl = (\ell_j)_{j=1}^\infty$ onto its spectrum $($represented by $\nu):$
\begin{equation}\label{7.17}
\eta := \sum_{j=1}^\infty \delta_{\{\ell_j^{-1} \}} \mapsto \nu = \nu(\eta) = \sum_{j, n=1}^\infty \delta_{\{n \cdot \ell_j^{-1} \},}
\end{equation}
where $\delta_{\{ y\}}$ denotes the Dirac point mass at $y > 0$ and the ``generalized fractal strings'' $\eta$ and $\nu$ are viewed as positive $($local$)$ measures $($or as tempered distributions$)$ on $(0, +\infty);$ see $[$Lap-vFr3, \S 6.3.2$]$. We also refer to $[$Lap-vFr3, \S 6.3.1$]$ for explicit formulas expressing $\eta$ and $\nu$ in terms of the underlying complex dimensions of $\mcl$, along with $[$Lap-vFr3, Chapter 4$]$ for the notion of {\em generalized fractal string}, based on the notion of local measure from $[$DolFr$]$, $[$JoLap$]$, $[$JoLapNie$]$ and $[$LapRa\u Zu1, Appendix $A].$
\end{rem}

\tab Viewed additively (that is, after having made the change of variable $x = e^t, \ t= \log x$, with $x >0$ and $t \in \mathbb{R}$), the (heuristic) spectral operator, denoted by $\mfa$, becomes the following map (where $f = f(t)$ is a suitable function of the new variable $t \in \mathbb{R}$):

\begin{equation}\label{7.18}
f(t) \mapsto \mfa(f)(t):= \sum_{n=1}^\infty f (t - \log n).
\end{equation}

\tab As before, let $\mathcal{P}$ denote the set of all prime numbers. Then, given any $p \in \mcp$, the associated ({\em local}) {\em Euler factor} $\mfa_p$ is given by the operator

\begin{equation}\label{7.19}
f(t) \mapsto \mfa_p(f) = \sum_{m=0}^\infty f(t - m \log p).
\end{equation}
The spectral operator $\mfa$ and its (operator-valued) Euler factors $\mfa_p$ (with $p \in \mcp$) are connected via the following (operator-valued) {\em Euler product} representation of $\mfa$:

\begin{equation}\label{7.20}
\mfa (f)(t) = \left(\prod_{p \in \mcp} \mfa_p (f)\right)(t),
\end{equation}
with $\mfa_p = (1-p^{-\partial})^{-1}$. Here, we have let $\partial = d/dt$ denote the differentiation operator (also called the {\em infinitesimal shift} of the real line). We have (for any $h \in \mathbb{R}$)

\begin{equation}\label{7.21}
(e^{-h \partial}) (f)(t) = f(t-h);
\end{equation}
in particular, we have

\begin{equation}\label{7.22}
(n^{-h})(f)(t) = f(t - \log n),
\end{equation}
so that we should also expect to be able to obtain the following (operator-valued) {\em Dirichlet series} representation of $\mfa$:

\begin{equation}\label{7.23}
\mfa (f)(t) = \left(\sum_{n=1}^\infty \ n^{-\partial} \right) (f)(t).
\end{equation}

\tab All of these semi-heuristic formulas and definitions are given without proper mathematical justification in [Lap-vFr3, \S 6.3.2] (and, originally, in [Lap-vFr2, \S 6.3.2] where the heuristic spectral operator $\mfa$ was first introduced). However, in [HerLap1--5], Hafedh Herichi and the author have developed a rigorous functional analytic framework within which (under suitable assumptions) all of these ``definitions'' and formulas are properly justified and, among many other results, the reformulation of the Riemann hypothesis obtained in \cite{LapMai2} (see Theorem \ref{T6.2} above) can be restated in operator theoretic terms; namely, as the ``quasi-invertibility'' of the spectral operator in all possible dimensions, except in the midfractal case $1/2$.\\

\tab We refer the interested reader to the forthcoming book \cite{HerLap1}, titled {\em Quantized Number Theory, Fractal Strings and the Riemann Hypothesis}, for a complete exposition of the theory. We limit ourselves here to a brief exposition and a few additional comments.\\

\tab For each fixed $c \in \mathbb{R}$, let $\mathbb{H}_c = L^2 (\mathbb{R}, e^{-2ct} 
dt)$,\footnote{Hence, $\mbh_c$ is the complex Hilbert space of (Lebesgue measurable, complex-valued) square integrable functions with respect to the absolutely continuous measure $e^{-2ct}dt$, and is equipped with the norm
\begin{equation}\label{7.24}
||f||_c := \left( \int_\mbr |f(t)|^2 \ e^{-2ct}\ dt \right)^{1/2} < \infty
\end{equation}
and the associated inner product, denoted by $(\cdot, \cdot)_c$.
}
and let $\partial = d/dt$ be the unbounded (and densely defined) operator acting on the Hilbert space $\mathbb{H}_c$ via $\partial (f) = f'$, the distributional (or else, the pointwise almost everywhere defined) derivative of $f$, for all $f$ in the domain $D (\partial)$ of $\partial$:

\begin{equation}\label{7.25}
D(\partial) := \{f \in C_{abs} (\mathbb{R}) \cap \mbh_c : f' \in \mbh_c \},
\end{equation}
where $C_{abs} (\mbr)$ is the space of (locally) absolutely continuous functions on $\mbr$ (see, e.g., \cite{Coh, Foll, Ru1}). Then, the {\em infinitesimal shift} $\partial = \partial_c$ is shown in [HerLap1--3] to be an unbounded normal operator on 
$\mbh_c$ (that is, $\partial^* \partial = \partial \partial^*$, where $\partial^*$ denotes the adjoint of the closed unbounded operator $\partial$; see, e.g., \cite{Kat} or \cite{Ru2}), with spectrum $\sigma (\partial) = \{\Res = c \}$. Furthermore, the associated group $\{e^{-h \partial}\} h \in \mbr $ is shown to be the {\em shift group} given (for all $f \in \mbh_c$) by \eqref{7.21} while $n^{- \partial}$ is therefore given by \eqref{7.22}. \\

\tab Moreover, for every $c > 1$, the quantized (i.e., operator-valued) Euler product and Dirichlet series representations \eqref{7.20} and \eqref{7.23} of $\mfa = \mfa_c$ hold, in the following strong sense: $\mfa = \sum_{n=1}^\infty n^{-\partial}$ and $\mfa = \Pi_{p \in \mathcal{P}} (1-p^{-\partial})^{-1}$, where both the series and the infinite product converge in $\mathcal{B} (\mbh_c)$, the Banach algebra of bounded linear operators on the Hilbert space $\mbh_c$. (See \cite{HerLap5} or [HerLap1, Chapter 7].) \\

\tab In addition, for all $f$ in a suitable dense subspace of $D (\partial)$ (and hence, of $\mbh_c$), the {\em spectral operator} $\mfa = \mfa_c$, now rigorously defined (for every $c \in \mbr$) by the formula $\mfa := \zeta (\partial)$, is given (for $c >0$) by an appropriate operator-valued version of the classic analytic continuation of the Riemann zeta function $\zeta = \zeta (s)$ to the open half-plane $\{\Res >0 \}$. Namely, formally, we have 
\begin{equation}\label{54.5}
\mfa = \zeta (\partial) = \frac{1}{\partial -1} + \sum_{n=1}^\infty \int_n^{n+1} (n^{- \partial} - t^{- \partial}) dt.
\end{equation}
(Compare, for example, with [Ser, \S VI.3] or the proof of Theorem \ref{T5.4} above given in \cite{LapPom2}.)\\

\tab Similarly, the {\em global spectral operator} $\mathcal{A} := \xi (\partial)$, where, as before, $\xi(s) := \pi^{-s/2} \Gamma (s/2)$ $\zeta (s)$ is the {\em global} (or {\em completed}) {\em Riemann zeta function}, is shown to be given by an operator-valued (or quantized) version of the standard analytic continuation of $\xi = \xi (s)$ to the entire complex plane (see, e.g., [Ti, \S 2.6] or [Lap6, \S 2.4]). Again, we refer the interested reader to \cite{HerLap5} or [HerLap1, Chapter 7] for the precise statements and many related results.\\

\tab We note that the spectral operator $\mfa = \mfa_c$, now defined for any $c \in \mbr$ by $\mfa := \zeta (\partial)$, is obtained via the functional calculus for unbounded normal operators (see \cite{Ru2}); so that, formally, one substitutes the infinitesimal shift $\partial$ for the complex variable $s$ in the usual definition of the Riemann zeta function (or, rather, in its meromorphic continuation to all of $\mbc$, still denoted by $\zeta = \zeta(s)$, for $s \in \mbc$).\\

\tab As it turns out, according to the celebrated spectral theorem (for unbounded normal operators, see \cite{Ru2}), what matters are the values of $\zeta = \zeta (s)$ along the spectrum $\sigma (\partial)$ of $\partial$; that is, according to a key result of [HerLap1,2], along

\begin{equation}\label{7.26}
\sigma (\partial) = c \ell (\zeta (\{s \in \mbc: \Res = c \}),
\end{equation} 
where $c \ell (G)$ denotes the closure of $G \subseteq \mbc$ in $\mbc.$\\

\tab We point out that for $c =1$, the unique (and simple) pole of $\zeta$, we must assume that $s \neq 1$ on the right-hand side of \eqref{7.26}. Alternatively, one views the meromorphic function $\zeta$ as a continuous function with values in the Riemann sphere $\widetilde{\mbc} := \mbc \cup \{\infty \}$, equipped with the standard chordal metric. (We set $\zeta (1) = \infty$.) Then, even for $c=1$, the extended spectrum $\widetilde{\sigma} (\mfa)$ of $\mfa$ is given by the right-hand side of \eqref{7.26}, {\em but without the closure}, where $\widetilde{\sigma} (\mfa) := \sigma (\mfa) \cup \{\infty \}$ if $\mfa$ is unbounded and $\widetilde{\sigma} (\mfa) := \sigma (\mfa)$ if $\mfa$ is bounded; that is, according to [HerLap1--5], if $c \leq 1$ or if $c >1$, respectively.\\

\tab One therefore sees how the properties of the Riemann zeta function $\zeta = \zeta(s)$ (for example, its range along the vertical line $\{\Res =1 \}$ or the existence of a pole at $s=1$) are reflected in the properties of the spectral operator $\mfa = \zeta(\partial)$ (for example, whether or not $\mfa$ is bounded or invertible or, equivalently, whether or not its spectrum $\sigma (\mfa)$ is compact or does not contain the origin, respectively).\\

\tab Formula \eqref{7.26} for $\sigma (\mfa)$ follows from a suitable version of the {\em spectral mapping theorem} (SMT) for unbounded normal operators (see [HerLap1, Appendix E]) according to which $\sigma (\zeta (\partial)) = c \ell (\zeta (\sigma(\partial)) \backslash \{1 \})$ or, equivalently, $\widetilde{\sigma} (\zeta (\partial)) = \zeta (\sigma (\partial))$ (where in the latter formula, $\zeta$ is viewed as a continuous $\widetilde{\mbc}$-valued function, as explained above). Note that for $c \neq 1,$ we have that $1 \notin \sigma (\partial)$, so that $\zeta = \zeta(s)$ is holomorphic and hence, continuous along $\sigma (\partial) = \{\Res = c \}$, whereas for $c =1$, $\zeta = \zeta (s)$ is meromorphic but has a singularity (more specifically, a simple pole) at $s =1 \in \sigma (\partial) $. Consequently, we can use the continuous (resp., meromorphic) version of SMT (stated and proved in [HerLap1, Appendix E]) in order to deduce that when $c \neq 1$ (resp., $c=1$), the identity \eqref{7.26} (resp., the counterpart of \eqref{7.26} when $c=1$) holds. \\

\tab Next, given $T >0$, one defines (still via the functional calculus for unbounded normal operators) a {\em truncated infinitesimal shift} $\partial^{(T)} = \partial _c^{(T)}$ (say, $\partial^{(T)} := \varphi^{(T)} (\partial)$, where $\varphi^{(T)}$ is a suitable cut-off function), so that (again by the aforementioned spectral mapping theorem, SMT), $\sigma (\partial^{(T)}) = [c -  iT, c + iT]$, where $c \in \mbr$ and $i := \sqrt{-1}$. (We refer to [HerLap1--3] for the precise definitions; see also Remark \ref{R17.5} just below.) Then, letting $\mfa^{(T)} := \zeta (\partial^{(T)})$, we obtain the {\em truncated spectral operator} $\mfa^{(T)}= \mfa_c^{(T)}$. We say that the spectral operator $\mfa$ is {\em quasi-invertible} if each of its truncations $\mfa^{(T)}$ is invertible (in the usual sense, for bounded operators), for every $T >0$. (Note that the normal operator $\partial^{(T)}$ is bounded since its spectrum is compact.)

\begin{rem}\label{R17.5}
For $c \neq 1$, the cut-off function $\varphi^{(T)}: \sigma (\partial) = \{\Res = c \} \rightarrow \mbc$ is assumed to satisfy the following two conditions$:$ $($i$)$ $\varphi^{(T)}$ is continuous and, in addition, $($ii$)$ $c \ell (\varphi^{(T)} (\{\Res =c \})) = [c -iT, c + iT])$. For $c=1$, we replace $($i$)$ by $($i'$):$ $\varphi^{(T)}$ has a $($necessarily unique$)$ meromorphic continuation to a connected open neighborhood of $\{\Res =1 \}$. $($Then, condition $($i'$)$ and $($ii$)$ imply that $($i$)$ also holds.$)$ Consequently, when $c \neq 1 $ $($resp., c =$1)$, one  can use the continuous $($resp., meromorphic$)$ version of the spectral mapping theorem $($SMT, see $[$HerLap1, Appendix E$])$ in order to deduce that 
\begin{align}\label{57.25}
\sigma (\partial^{(T)}) &= \sigma (\varphi^{(T)} (\partial))\\
\notag &= c \ell (\varphi^{(T)}(\{\Res =c \})) = [c- iT, c + iT], 
\end{align}
and especially, that 
\begin{align}\label{57.5}
\sigma(\mfa^{(T)}) &= \sigma (\zeta (\partial^{(T)}))\\
\notag &= c \ell (\zeta (\sigma (\partial^{(T)}))) = c \ell (\zeta ([c - iT, c + iT])),
\end{align}
where for $c = 1$, one should remove $\{ 1\}$ from $\sigma (\partial^{(T)}) = [c -iT, c + iT]$ in the last two equalities of \eqref{57.5}, while when $c \neq 1$, \eqref{57.5} takes the following simpler form (since then, $\zeta$ is continuous along the vertical line segment $[c - iT, c+ iT]):$
\begin{equation}\label{57.75}
\sigma (\mfa^{(T)}) = \zeta ([c -iT, c+ iT]).
\end{equation}
When $c=1$, \eqref{57.75} can be replaced by the following identity, which is equivalent to \eqref{57.5} interpreted as above $($since $\zeta (1) = \{\infty \} $ and $\zeta (s) \neq \infty$ for all $s \in \mbc, s \neq 1):$
\begin{equation}\label{57.8}
\widetilde{\sigma} (\mfa^{(T)}) := \sigma (\mfa^{(T)}) \cup \{\infty \} = \zeta ([1 -iT, 1+ iT]),
\end{equation}
where $($as in an earlier comment$)$ the meromorphic function $\zeta$ should be viewed as a $\widetilde{\mbc}$-valued continuous function in the right-hand side of \eqref{57.8} and where $\widetilde{\sigma} (\mfa^{(T)})$ denotes the extended spectrum of $\mfa^{(T)}$.
\end{rem}

\tab The full strength of the definition given in the following remark will be needed when we discuss Theorem \ref{T7.6} below. This definition and the accompanying property is also used in the statement and the proof of Theorem \ref{T7.3} (and hence, of Theorem \ref{T7.4} as well).

\begin{rem}\label{R7.3}
A possibly unbounded linear operator $L: D(L) \subseteq H \rightarrow H$ on a Hilbert space $H$, with domain $D(L)$, is said to be {\em invertible} if it is a bijection from $D(L)$ onto $H$ and if its inverse, $L^{-1}$, is bounded. $($If $L$ is closed, an assumption which is satisfied by all of the operators considered in the present subsection $($\S 7.4$)$, then $L^{-1}$ is automatically bounded, by the closed graph theorem $\cite{Kat, Foll, Ru1}.)$ Furthermore, essentially by definition of the spectrum, $L$ is invertible if and only if $0 \notin \sigma (L)$. $($See, e.g., $\cite{Kat, ReSi1, Ru2, Sc}.)$
\end{rem}

\tab We can now state the counterparts of Theorem \ref{T6.1} and Theorem \ref{T6.2} (the key results from \cite{LapMai2} discussed in \S6 above) in this context. (See Remark \ref{R7.5} below.)

\begin{theorem}[Analog of Theorem \ref{T6.1}, \cite{HerLap1,HerLap3}]\label{T7.3}
Given any $c \in \mbr$, the spectral operator $\mfa = \mfa_c$ is quasi-invertible if and only if $\zeta = \zeta (s)$ does not have any zeros along the vertical line $\{\Res = c\};$ i.e., if and only the ``$c$-partial Riemann hypothesis'' is true.
\end{theorem}

\tab Hence, much as in Corollary \ref{C6.3}, $\mfa_{1/2}$ is {\em not} quasi-invertible (since $\zeta$ has at least one zero along the critical line $\{\Res = 
1/2 \}$).\footnote{In fact, $\zeta$ has infinitely many zeros along the critical line, according to Hardy's theorem (see \cite{Edw, Ti}) but this is irrelevant here. We refer to [HerLap1,3] for a version of Theorem \ref{T7.3} for which this well-known fact actually matters.}
Furthermore, $\mfa_c$ is quasi-invertible for every 
$c > 1$. Actually, for $c >1$, $\mfa = \mfa_c$ is invertible (which implies that it is quasi-invertible) and its inverse is given by $\mfa^{-1} = \prod_{p \in \mcp} (1-p^{- \partial}),$ with the convergence of the infinite product holding in $\mathcal{B}(\mbh_c)$; see [HerLap1,5].\\

\tab Just as Theorem \ref{T6.2} is a consequence of Theorem \ref{T6.1}, the following key result follows from Theorem \ref{T7.3}.

\begin{theorem}[Analog of Theorem \ref{T6.2}, \cite{HerLap1, HerLap3}]\label{T7.4}
The spectral operator is quasi-invertible for every value of $c$ in $(0,1)$ other than in the midfractal case when $c = 1/2$ $($or, equivalently, for every $c \in (0, 1/2))$ if and only if the Riemann hypothesis is true.
\end{theorem}

\begin{rem}\label{R7.5}
Recall from \S6 that Theorems \ref{T6.1} and \ref{T6.2} are expressed in terms of the solvability of the inverse spectral problem $($ISP$)_D$ for all fractal strings of Minkowski dimension $D$. Here, the role played by the dimension $D$ is now played by the parameter $c$, while the solvability of $($ISP$)_D$ is now replaced by the quasi-invertibility of $\mfa_c$. In a precise sense, $c$ is the analog of $D$ in the present context; this analogy is based in part on some of the results of $[$LapPom2$]$ which are briefly discussed in \S3 above $($see Theorem \ref{T3.2} and the comments following it$)$, provided one thinks of the functions $f = f(t)$ as geometric counting functions of fractal strings (modulo the change of variable $x=e^{t}$).
\end{rem}

\tab We leave it to the interested reader to state (in the present context) the counterpart of Theorem \ref{T6.4} (about a mathematical phase transition at $c = 1/2$).\\

\tab Like Theorem \ref{T6.2}, Theorem \ref{T7.4} is a {\em symmetric criterion for the Riemann hypothesis} (RH). Indeed, in light of the functional equation \eqref{6.2} for $\zeta$ connecting $\zeta (s)$ and $\zeta (1-s)$, we can replace the open interval $(0,1/2)$ by the symmetric interval $(1/2,1)$, with respect to $1/2$. (See also Equation \eqref{27.5} above.) In contrast, the author has recently discovered the following {\em asymmetric criterion for} RH, as we now explain. (See \cite{Lap7}.)\\

\tab Let $\mfb = \mfb_c$ be the nonnegative self-adjoint operator defined by $\mfb := \mfa \mfa^* = \mfa^* \mfa$. Note that $\mfb$ is a nonnegative self-adjoint operator because $\mfa$ is normal (see, e.g., \cite{Kat}) and, like $\mfa$ itself, is unbounded for all $c \in (0,1)$, while it is bounded for all $c >1$ (see [HerLap1--3]). Then, $\mfb$ is invertible (in the sense of possibly unbounded operators, see Remark \ref{R7.3} above) if and only if $\mfb$ is bounded away from zero (i.e., if and only if there exists $\gamma > 0$ such that $(\mfb f, f)_c \geq \gamma ||f||_c^2$, for all $f \in D(\mfb)$). Furthermore, $\mfb$ is invertible if and only if $\mfa$ is invertible, which is the case if and only $0 \notin \sigma (\mfa)$ or, equivalently, $0 \notin \sigma (\mfb)$.

\begin{theorem}[An asymmetric reformulation of the Riemann hypothesis, \cite{Lap7}]\label{T7.6}
The following statements are equivalent$:$
\begin{enumerate}
\item[$($i$)$] The Riemann hypothesis is true.
\item[$($ii$)$] The spectral operator $\mfa$ is invertible for all $c \in (0,1/2).$
\item[$($iii$)$] The self-adjoint operator $\mfb$ is invertible for all $c \in (0, 1/2)$.
\item[$($iv$)$] For each $c \in (0, 1/2)$, $\mfb$ is bounded away from 
zero.
\end{enumerate}
\end{theorem}

\tab We know that in part ($iv$) of Theorem \ref{T7.6}, the implicit lower bound $\gamma = \gamma_c$ may vary with $c$ and even tend to $0$ as $c \rightarrow (1/2)^-$ or $c \rightarrow 0^+$.\\

\tab A priori, the phase transition observed at $c = 1/2$ in Theorem \ref{T7.6} just above is of a very different nature from the one observed in Theorem \ref{T6.2} (based on \cite{LapMai2}, see also Theorem \ref{T6.4}) or in Theorem \ref{T7.4} (based on [HerLap1,3], see the comment following Remark \ref{R7.5}). Indeed, under RH (and, in fact, if and only RH holds, by Theorem \ref{T7.6}), we have that, for all $c \in (0, 1/2)$, $\sigma (\mfa)$ is a closed unbounded subset of $\mbc$ not containing $0$ and hence, $\mfa$ is invertible. This follows from the expression \eqref{7.26} obtained in [HerLap1--3] for $\sigma (\mfa)$, combined with a conditional result (by Garunk\v stis and Steuding; see the proof of Lemma 4 and Proposition 5 in \cite{GarSte}) about the non-universality of $\zeta$ in the left critical strip $\{0 < \Res < 1/2 \}$.\\

\tab On the other hand, for any $c \in (1/2, 1)$, we have that $\sigma (\mfa) = \mbc$ and hence, that $\mfa$ is not invertible. This follows again from \eqref{7.26}, but now combined with the Bohr--Courant theorem \cite{BohCou} according to which for every $c \in (1/2, 1)$, the range of $\zeta$ along any vertical line $\{\Res = c \}$ is dense in $\mbc$. This latter fact is a consequence of the universality of $\zeta$ in the right critical strip $\{1/2 < \Res < 1 \}$, as was first observed and established by Voronin in [Vor1,2].\\

\tab Consequently, the (unconditional) universality of the Riemann zeta function $\zeta$ in the right critical strip $\{1/2 < \Res < 1 \}$ and the (conditional) non-universality of $\zeta$ in the left critical strip $\{0 < \Res < 1/2 \}$ are absolutely crucial in order to understand the mathematical phase transition occurring in the midfractal case when $c = 1/2$, according to Theorem \ref{T7.6}, if and only RH holds.\\

\tab We stress that the universality of $\zeta$, initially discovered by Voronin in \cite{Vor2} in the mid-1970s after he extended the Bohr--Courant theorem to jets of $\zeta$ (consisting of $\zeta$ and its derivatives), in \cite{Vor1}, has since been generalized in a number of ways; see, e.g., the books \cite{KarVor, Lau, Ste1}, along with [Bag1--2, Rei1, Her1, Her4, Ste2] and the many relevant references therein.\\

\tab Roughly speaking, the universality theorem, as generalized in [Bag1--2, Rei1], states that given any compact subset $K$ of the right critical strip $\{ 1/2 < \Res < 1 \}$ with connected complement in $\mbc$, every $\mbc$-valued continuous function on $K$ which is holomorphic and nowhere vanishing on the interior of $K$ can be uniformly approximated on $K$ by vertical translates of $\zeta$.\\

\tab The universality theorem has been extended to a large class of $L$-functions for which the generalized Riemann hypothesis (GRH) is expected to hold; see, e.g., \cite{Ste1} for an exposition, along with \cite{Ste2} and the relevant appendices of [HerLap1,4]. Accordingly, Theorem \ref{T7.6} can be extended to many $L$-functions. Similarly, but for different and simpler reasons, after a suitable modification in the definition of the truncated infinitesimal shift $\partial^{(T)}$, Theorems \ref{T7.3} and \ref{T7.4} (based on [HerLap1--4]) can be extended to essentially all of the $L$-functions (or arithmetic zeta functions) for which GRH is expected to hold (independently of whether or not the analog of the universality theorem holds for those zeta functions); see, e.g., [ParSh1,2], [Sarn], [Lap-vFr3, Appendix A] and [Lap6, Appendices B,C,E].\\

\tab Moreover, we point out that a counterpart in the present context (that is, for the spectral operator $\mfa = \zeta(\partial)$) of the universality of $\zeta$ is provided in [HerLap1,4]. Interestingly, the corresponding ``quantized universality'' then involves the family of truncated spectral operators $\{\mfa^{(T)} = \zeta (\partial^{(T)}) \}_{T > 0}$; that is, the complex variable $s$ is not replaced by the operator $\partial$, as one might naively expect, but by the family of truncated infinitesimal shifts $\{\partial^{(T)} \}_{T>0}$ discussed earlier in this subsection; see \cite{HerLap4} and \cite{HerLap1}.\\

\tab In addition, we note that the (conditional) phase transition at $c = 1/2$ associated with Theorem \ref{T7.6} (and also with Theorem \ref{T6.2}, from [LapMai1,2], which preceded the work in [BosCon1,2]; see also its interpretation given in Theorem \ref{T6.4}, from [Lap2,3]) is of a very different nature from the one studied by Bost and Connes in [BosCon1,2]. (See also [Con, \S V.11].) Indeed, the latter phase transition has nothing to do with the universality or with the critical zeros of $\zeta$ but instead, is merely connected with the pole of $\zeta(s)$ at 
$s=1$.\footnote{A similar phase transition (or ``symmetry breaking'') is observed at $c=1$, as is discussed in [HerLap1--4], since $\mfa$ is bounded and invertible for $c >1$, unbounded and not invertible for $1/2 \leq c \leq 1$, while, likewise, $\sigma (\mfa)$ is a compact subset of $\mbc$ not containing the origin for $c > 1$ and $\sigma (\mfa) = \mbc$ is unbounded and contains the origin if $1/2 < c < 1$.}
However, it is natural to wonder whether the conditional phase transition occurring at $c = 1/2$ in Theorem \ref{T7.6} (if and only if RH holds) can be interpreted physically and mathematically (as in [BosCon1,2]) as some kind of ``symmetry breaking'' associated with a suitable physical model (in quantum statistical physics or quantum field theory, for example) as well as corresponding to a change in the nature of an appropriate symmetry group yet to be attached to the present situation.

\begin{rem}\label{R7.7}
It is also natural to wonder what is the inverse of the spectral operator, when $0 < c < 1/2$ and assuming that RH holds. Naturally, according to the functional calculus, we must have $\mfa^{-1} = (1/\zeta)(\partial)$, even if $\mfa^{-1}$ is not bounded $($i.e., even if $D (\mfa^{-1}) \neq \mbh_c$ or equivalently, if $R(\mfa) \neq \mbh_c$, where $R(\mfa)$ denotes the range of $\mfa)$. We conjecture that under RH, we have for all $c \in (0,1/2)$ and all $f \in D(\mfa^{-1}) = \mbh_c, \ \mfa^{-1} (f) = \sum_{n=1}^\infty \mu (n) n^{-\partial} (f)$, so that $\mfa^{-1} (f)(t) = \sum_{n=1}^\infty \mu (n) f (t - \log n)$, both for a.e. $t \in \mbr$ and as an identity between functions in $\mbh_c$. $($Here, $\mu = \mu (n)$ denotes the classic M\" obius function, given by $\mu (n) = 1$ or $-1$, respectively, if $n \geq 2$ is a square-free integer that is a product of an even or odd number of distinct primes, respectively, and $\mu (n)=0$ otherwise; see, e.g., $\cite{Edw}$ or $\cite{Ti}.)$ Observe that if that were the case, then ``quantized Dirichlet series'' would behave very differently from ordinary $($complex-valued$)$ Dirichlet series. In particular, the quantized Riemann zeta function $\mfa = \zeta (\partial)$ would be rather different from its classic counterpart, the ordinary Riemann zeta function, $\zeta = \zeta (s)$ $($where $s \in \mbc)$. Indeed, as is well known, even if the Riemann hypothesis is true, the series $\sum_{n=1}^\infty \mu (n) \ n^{-s}$ cannot converge for any value of $s := s_0$ with $0 < Re(s_0) < 1/2$. Otherwise, it would follow from the standard properties of $($numerical$)$ Dirichlet series and the M\" obius inversion formula $($see, e.g., $\cite{Edw})$ that the sum of the series $\sum_{n=1}^\infty \mu (n) n^{-s}$ would have to be holomorphic and to coincide with $(1/\zeta) (s)$ in the half-plane $\{ \Res > \text{Re}(s_0) \}$, which is, of course, impossible because the meromorphic function $1/ \zeta$ must have a pole at every zero of $\zeta$ along the critical line $\{\Res = 1/2 \}$ $($of which there are infinitely many$)$.
\end{rem}
\tab We close this discussion by a final comment related to the operator $\mfb$ $($appearing in the statement of parts ($iii$) and ($iv$) of Theorem \ref{T7.6}$)$ and to the possible origin of the $($conditional$)$ phase transition occurring in the midfractal case $c=1/2$.

\begin{rem}\label{R7.8}
The nonnegative self-adjoint operator $\mfb = \mfa \mfa^* = \mfa^* \mfa$ is given by the following formula$:$
\begin{equation}\label{7.27}
\mfb(f)(t) = \zeta (2c) \sum_{(k,n) =1}f \left(t - \log \frac{k}{n} \right) n^{-2c}, 
\end{equation}
where the sum is taken over all pairs of integers $k,n \geq 1$ without common factor. The appearance of the terms $n^{-2c}$ $($for $n \geq 1)$ and of the factor $\zeta (2c)$ are very interesting features of this formula $($which first appeared in $[$Lap-vFr2--3, \S 6.3.2$]$ without formal justification$)$. A priori, the above formula was derived $($in $[$HerLap1,5$]$, motivated in part by the new result of $\cite{Lap7}$ described in Theorem \ref{T7.6} above$)$ by assuming that $c > 1$. However, the series $\sum_{n=1}^\infty \ n^{-2c}$ is convergent for $c > 1/2$ and the term $\zeta (2c)$ is singular only at $c = 1/2$, due to the pole of $\zeta (s)$ at $s=1$. This may shed new light on the origin of the phase transition occurring at $c =1/2$ in the statements of Theorem \ref{T7.6}. \\
\indent \indent Given the reformulation of the Riemann hypothesis provided above in terms of the invertibility of the spectral operator $\mfb = \mfb_c$, where $\mfb = \mfa^* \mfa = \mfa \mfa^*$ $($see the equivalence of $($i$)$, $($iii$)$ and $($iv$)$ in Theorem \ref{T7.6}$)$, it is tempting to extend formula \eqref{7.27} to a suitable class $\mcc$ of ``test functions'' $f = f(t)$ $($technically, a ``core'' for the operator $\mfb$, that is, a dense subspace of $D(\mfb)$ in the Hilbert graph norm $|||f|||_c := (||f||_c^2 + ||bf||_c^2)^{1/2}$) so that it becomes valid for all $c \in (0, 1/2)$. More specifically, for each $c \in (0,1/2)$, we would like to show that there exists a constant $\gamma = \gamma_c >0$ $($which may depend on the parameter $c)$ such that 

\begin{equation}\label{7.28}
||bf||_c \geq \gamma ||f||_c,
\end{equation}
for all $f \in 
\mcc$. $($Compare with part $($iv$)$ of Theorem \ref{T7.6} above as well as with the comments preceding the statement of Theorem \ref{T7.6}.$)$ \\
\indent \indent In light of the new asymmetric criterion for RH obtained in $\cite{Lap7}$ $($see Theorem \ref{T7.6} and, especially, the equivalence of $($i$)$ and $($iv$)$ in Theorem \ref{T7.6}$)$, this conjectured inequality \eqref{7.28} would imply $($in fact, would be equivalent to$)$ the Riemann hypothesis. Thus far, however, the author has only been able to verify it for a certain class of test functions, unfortunately not yet large enough to fulfill the required conditions.
\end{rem}

\tab A last comment is in order.

\begin{rem}\label{R7.9}
In two works in preparation $(\cite{Lap8}$ and $\cite{Lap9})$, the author explores some of the applications of this new formalism $($``quantized number theory'' from $[$HerLap1--5$]$ and $[$Lap7$])$ to potentially reformulating the Weyl conjectures $($for curves and higher-dimensional varieties over finite fields or, equivalently, for function fields$)$,\footnote{See, e.g., [Lap6, Appendix B] and the references therein.} 
as well as for constructing generalized Polya--Hilbert 
operators\footnote{See, e.g., [Lap6, Chapters 4 and 5], along with the relevant references therein.}
with spectra the Riemann zeros or, in fact, the zeros $($and the poles$)$ of general $L$-functions. For this purpose, the author has introduced a related but different formalism, in $[$Lap8,9$]$ associated with harmonic analysis and operator theory in weighted Bergman spaces of analytic functions (see, e.g., $\cite{HedKorZh}, 
\cite{AtzBri})$.\footnote{Some of this work may eventually become joint work with one of the author's current Ph.D. students, Tim Cobler.}
On the one hand, this new functional analytic framework offers greater flexibility and ease of use, since it only involves bounded operators, albeit of an unusual nature. On the other hand, the framework $($from $[$HerLap1--5, Lap7$])$ discussed in this section presents the significant advantage, in particular, of naturally parametrizing the critical strip $0 < \Res < 1$ by means of the dimension parameter $c \in (0,1)$. Only future research on both approaches will help us to eventually determine whether one formalism should be preferred to the other one, or as is more likely to happen, whether both formalisms should be used in order to further develop different aspects of quantized number theory.
\end{rem}



\begin{acknowledgement}
This research was partially supported by the US National Science Foundation (NSF) under the grants DMS-0707524 and DMS-1107750 (as well as by many earlier NSF grants since the mid-1980s). Part of this work was completed during several stays of the author as a visiting professor at the Institut des Hautes Etudes Scientifiques (IHES) in Paris/Bures-sur-Yvette, France.
\end{acknowledgement}

\end{document}